\documentclass[usenatbib,fleqn]{mnras}
\usepackage{graphicx,hyperref,amsmath,amssymb,xspace}
\usepackage[dvipsnames]{xcolor}

\title{The last breath of the Sagittarius dSph}

\author[E. Vasiliev \& V. Belokurov]{
Eugene Vasiliev$^{1,2}$\thanks{E-mail: eugvas@lpi.ru} and Vasily Belokurov$^1$\\
$^1$Institute of Astronomy, Madingley road, Cambridge, CB3 0HA, UK\\
$^2$Lebedev Physical Institute, Leninsky prospekt 53, Moscow, 119991, Russia}

\newcommand{\Gaia}{\textit{Gaia}\xspace}

\newcommand{\kms}{km\:s$^{-1}$\xspace}
\newcommand{\masyr}{mas\:yr$^{-1}$\xspace}

\begin{document}
\date{Accepted 2020 July 15. Received 2020 July 14; in original form 2020 June 9}
\pagerange{4162--4182}\volume{497}\pubyear{2020}
\setcounter{page}{4162}
\maketitle

\begin{abstract}
We use the astrometric and photometric data from \Gaia Data Release 2 and line-of-sight velocities from various other surveys to study the 3d structure and kinematics of the Sagittarius dwarf galaxy.
The combination of photometric and astrometric data makes it possible to obtain a very clean separation of Sgr member stars from the Milky Way foreground; our final catalogue contains $2.6\times10^5$ candidate members with magnitudes $G<18$, more than half of them being red clump stars. We construct and analyze maps of the mean proper motion and its dispersion over the region $\sim30\times12$ degrees, which show a number of interesting features. The intrinsic 3d density distribution (orientation, thickness) is strongly constrained by kinematics; we find that the remnant is a prolate structure with the major axis pointing at $\sim45^\circ$ from the orbital velocity and extending up to $\sim 5$~kpc, where it transitions into the stream. We perform a large suite of $N$-body simulations of a disrupting Sgr galaxy as it orbits the Milky Way over the past 2.5~Gyr, which are tailored to reproduce the observed properties of the remnant (not the stream). 
The richness of available constraints means that only a narrow range of parameters produce a final state consistent with observations. The total mass of the remnant is $\sim4\times10^8\,M_\odot$, of which roughly a quarter resides in stars. The galaxy is significantly out of equilibrium, and even its central density is below the limit required to withstand tidal forces.
We conclude that the Sgr galaxy will likely be disrupted over the next Gyr.
\end{abstract}


\section{Introduction}

Sagittarius dwarf galaxy (Sgr) is one of the closest and most massive satellites of the Milky Way (MW), however, its structure and properties are still not well known. By far the most spectacular feature of this galaxy is the giant tidal stream covering a large fraction of the sky, indicating an ongoing disruption of the satellite. The core of Sgr galaxy itself is located behind the Galactic bulge, and was discovered only relatively recently \citep{Ibata1994,Ibata1995}. Due to the small distance from the Galactic centre ($\lesssim 20$~kpc) and a distorted shape, it was immediately suspected to be tidally disrupting, and indeed a few years later the full extent of the tidal stream was uncovered in the 2MASS survey \citep{Majewski2003}. Since the stream spans a large range of galactocentric radii and wraps around the Galaxy more than once, it has been studied extensively as a probe of the Galactic gravitational potential \citep[e.g.,][]{Law2010}. On the other hand, the dynamical state and properties of the Sgr remnant did not receive a comparable level of attention.

\citet{Penarrubia2010} developed a scenario in which the Sgr galaxy had an initially rapidly rotating stellar disc, which explained the bifurcation seen in the Sgr stream \citep{Belokurov2006}. However, subsequent observation of the line-of-sight velocity field across the remnant \citep{Penarrubia2011} did not agree with the predictions of that scenario. At the same time, \citet{Lokas2010} presented another model for the Sgr remnant, which also started off as a rotating disc galaxy, but developed a strong bar induced by tidal torques from the MW, nearly eliminating all rotation in the remnant. This model was in a better agreement with the then-available line-of-sight kinematics \citep{Frinchaboy2012}, and remains the most recent model of the Sgr core.

The original mass of the Sgr progenitor and the current mass of the remnant have continued to be a subject of debate since the discovery of the dwarf. It became apparent sufficiently early on that a satellite as a large as Sgr coming as close to the Galactic disc could impart plenty of damage \citep[see e.g.,][]{Ibata1998,Bailin2003}. While these earlier studies assumed a dwarf with a mass of order of $10^9$ M$_{\odot}$, the subsequent census of the stellar content of the satellite revealed that its total mass could be as high as $10^{11}$ M$_{\odot}$ \citep[see][]{NiedersteOstholt2010}. This heavier Sgr appeared a much more serious threat to the integrity of the Galaxy: now it could make the MW ring, seed spiral waves \citep[see][]{Purcell2011} or even disrupt the stellar disc altogether \citep[see][]{Laporte2018}.

The second data release (DR2) of the \Gaia space observatory \citep{Brown2018} dramatically expanded our knowledge of stellar kinematics in the MW and revealed that the Galactic disc is presently out of equilibrium, strongly implying a recent interaction with a massive perturber, quite likely the Sgr dwarf \citep[see e.g.][]{Antoja2018, Carrillo2019, Laporte2019, BlandHawthorn2019}. It remains to be established, however, whether these numerical models of the Sgr--MW interaction satisfy all of the observational constraints on the dwarf's mass and its loss rate. While several studies used \Gaia DR2 data to refine the all-sky view of the Sgr stream \citep{Antoja2020, Ibata2020, Ramos2020}, no detailed analysis of the stellar kinematics in the Sgr remnant existed -- an omission we aim to address in the present paper. This will pave the way to establishing the current dynamical state of the remnant, its bound mass and will help us constrain the Sgr in-fall conditions.

We first describe the procedure for selecting candidate Sgr members from the entire \Gaia catalogue in Section~\ref{sec:membership}, which simultaneously provides the kinematic maps of the mean proper motion (PM) and its dispersion across the galaxy. In Section~\ref{sec:vlos} we augment these data with the line-of-sight kinematics derived from several existing datasets. In Section~\ref{sec:photometry} we discuss the 3d structure of the Sgr remnant inferred from photometry, and estimate its stellar mass by examining the distribution of stars in absolute magnitudes. We analyze the observed kinematic properties of the remnant in Section~\ref{sec:kinematics}, and compare them with tailored $N$-body models of a disrupting satellite in Section~\ref{sec:models}. Finally, Section~\ref{sec:summary} wraps up. 

\section{Membership selection}  \label{sec:membership}

\begin{figure}
\includegraphics{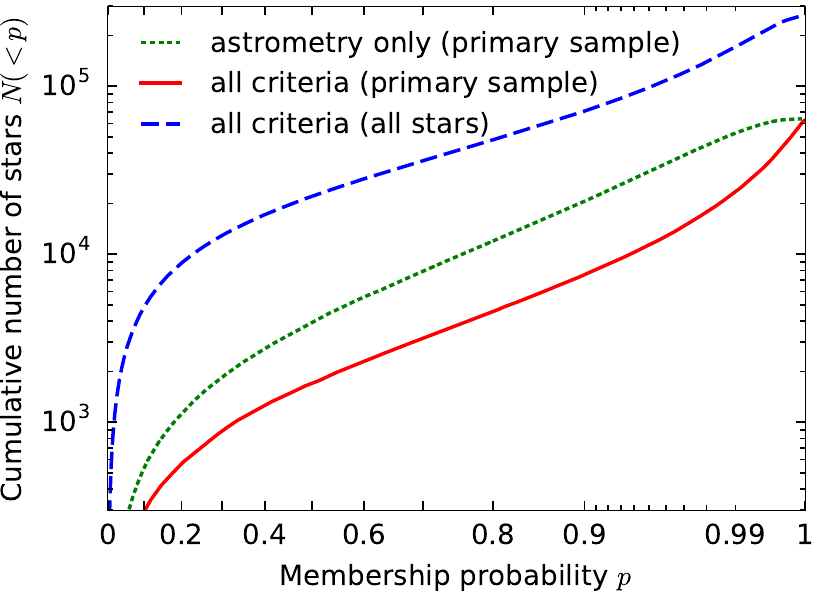}
\caption{Distribution of membership probabilities according to various criteria: number of member stars with membership probability not exceeding the value on the abscissa. Dotted green -- primary sample ($2.5\times10^6$ stars, $\sim 6\times10^4$ members), using astrometric information only (parallaxes and PM); solid red -- the same primary sample but using all available criteria (astrometry, CMD and location); dashed blue -- entire sample ($10^7$ stars, $2.5\times10^5$ members). The use of multiple classification criteria results in a very sharp distinction between members and field stars: half of member stars in the primary sample have probability of membership $p_\mathrm{(memb)}\gtrsim 99.5\%$, and 80\% -- $p_\mathrm{(memb)}\gtrsim 95\%$, and among all stars, 2/3 have $p_\mathrm{(memb)}\gtrsim 95\%$.}
\label{fig:membership}
\end{figure}

The centre of the Sgr galaxy coincides with the globular cluster M~54 (NGC~6715) and has ICRS coordinates $\alpha_0=283.764^\circ$, $\delta_0=-30.480^\circ$ or Galactic coordinates $l_0=5.607^\circ$, $b_0=-14.087^\circ$. Its major axis is approximately parallel to the $b$ axis of the Galactic coordinate system or the $\alpha$ axis of the ICRS coordinate system. We consider a hexagonal region defined by a combination of Galactic coordinates $0^{\circ}\le l \le 15^{\circ}, -35^{\circ} \le b \le -6^{\circ}$ and Declination $-38^{\circ}<\delta<-25^{\circ}$. The upper limit on $b$ is dictated by the rapidly increasing density of Milky Way foreground stars as one approaches the Galactic plane ($b=0$), while the lower limit allows us to include the important transition from the Sgr remnant to the stream, which occurs around $\sim 15^\circ$ from the Sgr centre, as will be shown later. The boundaries on $l$ and $\delta$ together select a strip of width $10-13^\circ$ along the minor axis, which encloses virtually all possible Sgr members.

We use a multi-stage procedure to select candidate member stars and to determine their astrometric parameters.  
The entire parent sample contains $1.4\times 10^7$ stars with $G<19$ in the region defined above, which have astrometric and colour information in \Gaia. We compute the extinction-corrected $G$-band magnitudes ($G_0$) and colours ($G_\mathrm{BP,0}-G_\mathrm{RP,0}$) as per Equation~1 in \citet{Babusiaux2018}. In the subsequent analysis, we retain all stars with $13\le G_0 \le 18$ ($\sim10^7$). This magnitude range includes blue horizontal branch (BHB), red clump (RC) and RR Lyrae stars in Sgr dSph. Note however that the uncertainties of these stars' PM measurements are relatively large and they overlap strongly wit the MW foreground population in the color-magnitude diagram (CMD). On the other hand, Sgr red giant (RG) stars with $G_0<17$ have smaller PM errors and are redder than most of the MW stars in this magnitude range, making the classification more reliable.  We therefore define the ``primary'' subset of $\sim 2.5\times 10^6$ stars, which is used in the initial classification and determination of the PM field of the dwarf. These stars must satisfy $13\le G_0 \le 17$, $|\varpi| \le 5\times \epsilon_\varpi$ and a number of additional quality filters recommended by \citet{Lindegren2018}:
\texttt{astrometric\_excess\_noise}${}<1$, \texttt{RUWE}${}<1.3$,
\texttt{phot\_bp\_rp\_excess\_factor}${}<1.3 + 0.06\,
\mathtt{bp\_rp}^2$.

We build a multidimensional mixture model forthe two components: Sgr (object) and MW (foreground or field) stars, writing the distribution function of $c$-th component in terms of the following parameters: position on the sky plane ($\alpha,\delta$), parallax and PM ($\varpi, \mu_\alpha, \mu_\delta$), position in the \Gaia colour-magnitude diagram (CMD) $G_0, G_\mathrm{BP,0}-G_\mathrm{RP,0}$, and -- for a small fraction of stars -- additional criteria such as the line-of-sight velocity $v_\mathrm{LOS}$. The distribution function of $c$-th component in each of these subspaces is denoted as $p_{(c)}^\mathrm{(loc)}$ (location), $p_{(c)}^\mathrm{(ast)}$ (astrometry), $p_{(c)}^\mathrm{(CMD)}$ (photometry), $p_{(c)}^\mathrm{(vel)}$ (line-of-sight velocity). These probability distributions are normalized so that the integral of $p_{(c)}^{(\dots\!)}$ over its respective subspace is unity.

\begin{figure*}
\includegraphics{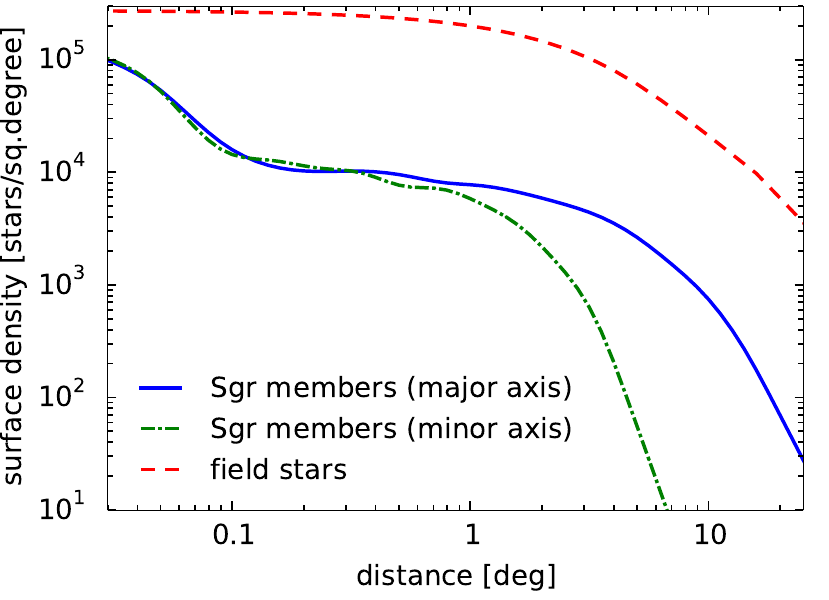} \hspace{18pt} \includegraphics{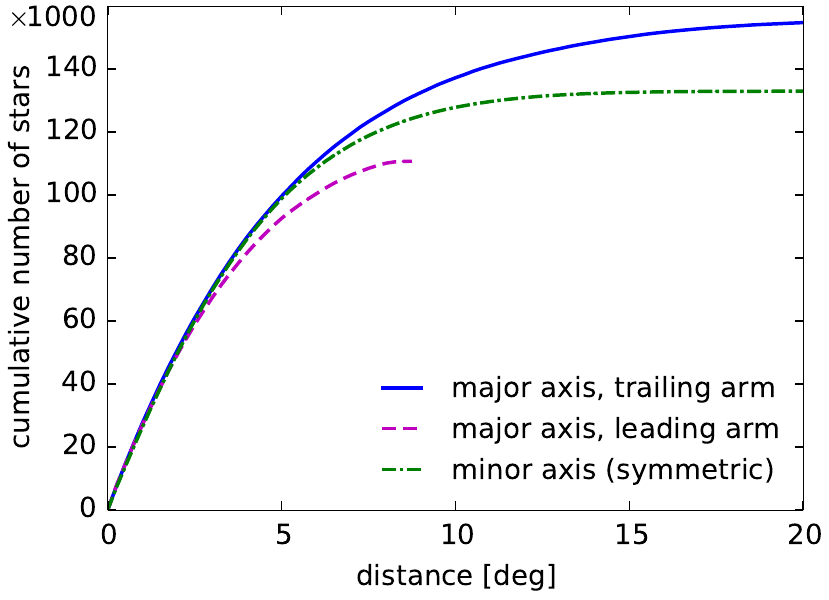}
\caption{
\textit{Left panel:} surface density profiles of Sgr members and field stars as functions of distance. For Sgr members, we plot separately the profiles along the major (solid blue) and the minor (dot-dashed green) axes as functions of the distance from the dwarf's centre. For field stars (dashed red), the abscissa is the offset in Galactic latitude $b$ from the region boundary at $b=-6$. The density of Sgr members (except for the central bump, corresponding to the globular cluster M~54) is well described by a King profile with axis ratio $\sim1:3$, core radius along the major axis $4^\circ$, and King parameter $W_0=4$, which has a tidal radius of $\sim 25^\circ$.\protect\\
\textit{Right panel:} cumulative number of Sgr members as a function of distance. Solid blue and dashed magenta lines show the profiles along the major axis in the direction of the trailing and leadin arm, respectively. The latter one is truncated at $\sim 8^\circ$ since the footprint of our catalogue ends there, but it already shows signs of incompleteness at smaller distances. Dot-dashed green line show the profile along the minor axis (symmetrized and divided by two) as a function of distance multiplied by the axis ratio 2.8.
} \label{fig:density_profile}
\end{figure*}

In a given subset of stars (e.g., a specific region on the sky), the fraction of Sgr members $\eta$ can be derived from the following argument. Consider a star with observed parameters $X\equiv \{X^\mathrm{(loc)}, X^\mathrm{(ast)}, X^\mathrm{(CMD)}, \dots\}$ The likelihoods of finding such a star among Sgr members or among field stars are, respectively,
\begin{equation}  \label{eq:likelihood_member_field}
\begin{array}{ll}
\mathcal L_{\mathrm{(memb)}, i} \!\!\!&= \eta\,p_\mathrm{(memb)}(X_i) , \\
\mathcal L_{\mathrm{(field)}, i}\!\!\!&= (1-\eta)\,p_\mathrm{(field)}(X_i) ,
\end{array}
\end{equation}
where $p_{(c)}(X_i)$ is the product of distribution functions in each subspace $p_{(c)}^\mathrm{(ast)}(X_i) \times p_{(c)}^\mathrm{(CMD)}(X_i) \times \dots$, and the probability of this star to be a Sgr member is
\begin{equation}  \label{eq:membership_probability}
p_{\mathrm{(memb)}, i} = \frac{\mathcal L_{\mathrm{(memb)}, i}}
{\mathcal L_{\mathrm{(memb)}, i} + \mathcal L_{\mathrm{(field)}, i}} .
\end{equation}
On the other hand, the membership fraction of the entire sample is
\begin{equation}  \label{eq:membership_fraction}
\eta = \frac{1}{N} \sum_{i=1}^N p_{\mathrm{(memb)}, i} .
\end{equation}
It can be determined iteratively, by repeating the steps (\ref{eq:likelihood_member_field}--\ref{eq:membership_fraction}) until the estimate of $\eta$ converges.
On the other hand, this iterative procedure may be also used to update the probability distributions of both member and field stars. Namely, at each iteration, one recomputes the parameters of $p_{(c)}^{(\dots)}$ from the values of $X_{i} |_{i=1}^N$, weighted by the current estimates of membership probability of each star. This approach is known as the expectation/maximization (EM) algorithm \citep[e.g.,][]{Dempster1977}.

We follow this strategy, splitting it into several stages, in which the distribution functions in different subspaces are updated one by one.  At the first stage, we determine the distribution functions of member and field stars in the 3-dimensional astrometric subspace (parallax $\varpi$ and PM $\mu_\alpha, \mu_\delta$). These are represented as a 5-component Gaussian mixture model, with the narrowest component being the Sgr stars, and the remaining ones represent the foreground population. We use the Extreme Deconvolution method \citep{Bovy2011}, which takes into account the observational errors of each star and constructs the intrinsic distribution function, which is broadened by measurement errors before evaluating the likelihoods (\ref{eq:likelihood_member_field}). We increase the parallax and PM errors quoted in the \Gaia catalogue by $10\%$ to compensate a slight underestimate of formal errors, as discussed by \citet{Lindegren2018}.  Since the PM distributions of both the Sgr and the MW stars vary considerably across the entire region of the sky, we perform this analysis separately in 8 partially overlapping macro-regions on the sky (4 strips in $b$ and 2 strips in $l$). Already at this stage it became clear that one needed to take into account the spatial gradients in the PM of the Sgr stars. We determined the overall trends in the mean PM of Sgr members across the entire region from the preliminary round of analysis, and subtracted them from the PM of all stars before running the mixture model. The following expressions approximate the mean PM values with an r.m.s.\ scatter of only $0.02$~\masyr across the entire region:
\begin{equation}
\begin{array}{rll}
\overline{\mu_\alpha} &\!\!\!=&\!\!\! -2.69 + 0.009\,\Delta\alpha - 0.002\,\Delta\delta - 0.00002\,\Delta\alpha^3, \\
\overline{\mu_\delta} &\!\!\!=&\!\!\! -1.35 - 0.024\,\Delta\alpha - 0.019\,\Delta\delta - 0.00002\,\Delta\alpha^3, 
\end{array}
\end{equation}
where $\Delta\alpha \equiv \alpha-\alpha_0, \Delta\delta \equiv \delta-\delta_0$ are the offsets in degrees from the Sgr centre.

\begin{figure*}
\includegraphics{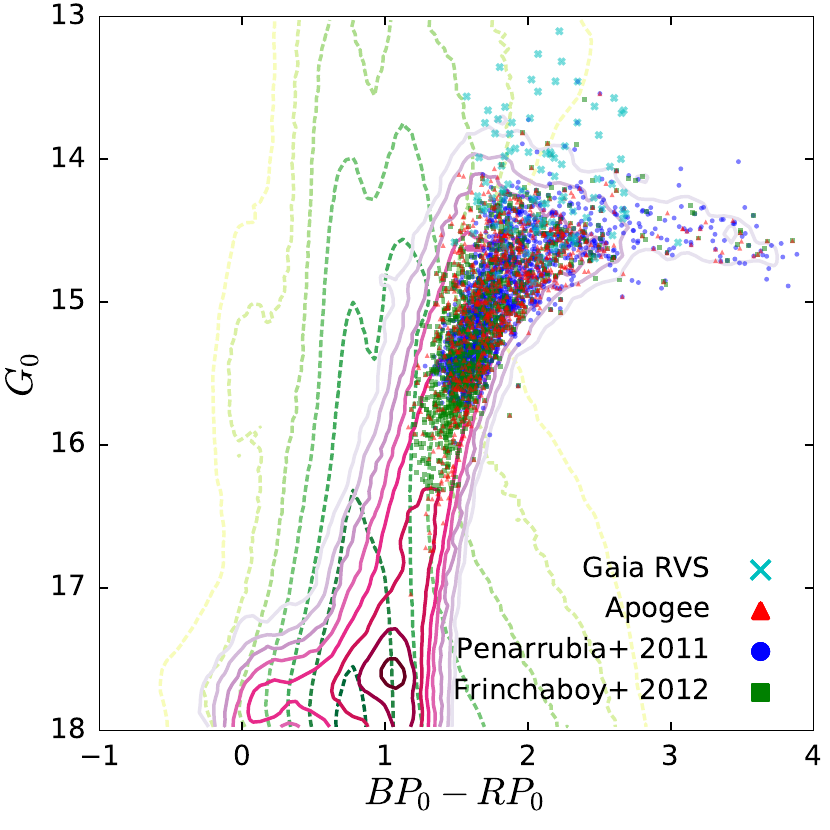} \hspace{18pt} \includegraphics{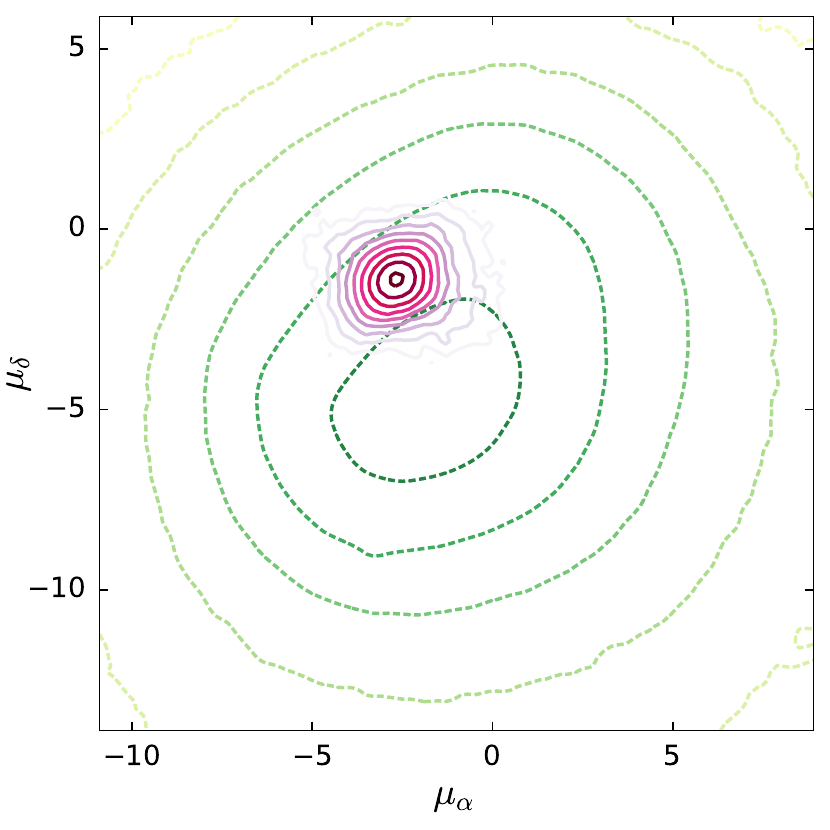}
\caption{
\textit{Left panel:} extinction-corrected CMDs of Sgr members (solid contour lines) and MW field stars (dashed), as inferred from the combination of all classification criteria. Points show the stars with measured line-of-sight velocities from various datasets. Contours are separated by 0.5 dex. The bright part of the Sgr red giant branch has virtually no overlap with the MW stars, which was used in previous studies to select Sgr members without relying on astrometric information. However, for the majority of Sgr stars in the final catalogue, photometry alone would not be sufficient to determine membership, although it does help to sharpen the classification.\protect\\
\textit{Right panel:} distribution of Sgr members (solid contours) and MW field stars (dashed) in the PM space; contours separated by 0.5 dex.
} \label{fig:CMD}
\end{figure*}

The first step (based on astrometry alone) already allows one to build a representative sample of Sgr members (Figure~\ref{fig:membership}, dotted green line). In the next step, this astrometric classification is used to determine the spatial distribution and construct colour--magnitude diagrams (CMDs) of both populations in the two-dimensional space of extinction-corrected apparent magnitudes $G_0$ vs. $G_\mathrm{BP,0}-G_\mathrm{RP,0}$. The spatial variation of the stellar density of the MW field stars is assumed to depend on the Galactic latitude $b$ only, while the density of the Sgr member stars is assumed to be a function of ellipsoidal radius $\tilde R \equiv \sqrt{X^2 + (Y/q)^2}$, where $X$ and $Y$ are the coordinates along the major and minor axes, respectively, and $q$ is the flattening parameter. Both density profiles are represented as free-form cubic splines with $\sim 10$ nodes each. We account for the fact that the footprint of our catalogue is truncated at $\sim 8^\circ$ from the Sgr centre in the direction of the leading arm by doubling the contribution of all stars at distances $>8^\circ$ in the trailing arm to the density profile of members. The CMD distribution functions are represented as 2d histograms in the range $13<G_0<18,\; -1<G_{BP,0}-G_{RP,0}<4$; we use 0.05-mag bins and additionally smooth the histograms with a 0.1-mag Gaussian kernel. As the photometric errors are relatively small and depend mainly on the $G$-band magnitude, the CMDs are constructed for the error-broadened, not intrinsic distributions (unlike the astrometric distributions). We follow the iterative EM procedure outlined above. At each iteration, the CMD histograms, the values of spline functions and the parameter $q$ are recomputed, using the current membership probability estimates $p_{\mathrm{(memb)},i}$ of each star as weight factors. Then these probability estimates are updated using the improved distribution functions of both populations (Equations~\ref{eq:likelihood_member_field}, \ref{eq:membership_probability}), and finally the overall fraction of the Sgr members $\eta$ is updated in Equation~\ref{eq:membership_fraction}. Initially, we determine density profiles and global CMD for both populations in the entire region, and then proceed by building more localized CMDs of field stars in separate macro-regions, while keeping the overall density profiles and the Sgr CMD fixed. The resulting density profiles, CMDs and PM distributions are shown in Figures~\ref{fig:density_profile}, \ref{fig:CMD}.

After constructing the CMD distribution functions, we return to estimating the parameters of the astrometric distribution function $p_\mathrm{(memb)}^\mathrm{(ast)}$ of Sgr members, but now determine the mean PM and its dispersion tensor in smaller spatial regions. We divide the entire region into 200 polygonal bins containing roughly equal numbers of member stars, using the Voronoi binning scheme of \citet{Cappellari2003} with some manual postprocessing. For all stars in each region, we again run the EM procedure, at each iteration recomputing the 5 parameters of the Gaussian distribution $p_\mathrm{(memb)}^\mathrm{(ast)}$: mean $\mu_\alpha,\mu_\delta$ and the components of symmetric covariance matrix. At this stage, we use the CMD distributions of both populations determined previously, but not the distance distributions (since these are normalized to unity in the entire region, not in each bin). We also keep fixed the parameters of the field distribution $p_\mathrm{(field)}^\mathrm{(ast)}$ (recall that they are also spatially-dependent, but vary on a larger scale than the bin sizes). We use only stars in the primary sample to determine the parameters of $p_\mathrm{(memb)}^\mathrm{(ast)}$, but then compute the membership probabilities for all stars. This avoids a possible inflation of the PM dispersion by fainter stars with lower PM precisions, and more importantly, by stars with unreliable astrometry, which did not pass the quality filters. We estimated the uncertainties on the mean PM (which are dominated by the spatially correlated systematic errors in \Gaia astrometry), using the method detailed in \citet{Vasiliev2019b}; these are $\lesssim 0.05$~\masyr. The statistical uncertainties on the PM dispersions are $\lesssim0.01$~\masyr. The resulting kinematic maps (mean PM and its dispersion) are discussed in Section~\ref{sec:kinematics}, and we provide the derived values in Table~\ref{tab:pm}.

Finally, we again run the EM procedure on the entire sample, keeping fixed all distributions except the density profile, to obtain the membership probability for each star (\ref{eq:membership_probability}). Figure~\ref{fig:membership} demonstrates that the combination of all selection criteria produces a very sharp distinction between members and field stars (almost 90\% of candidate member stars among the primary sample have membership probability exceeding 90\%, while this fraction drops to 75\% for the entire catalogue). The right panel of Figure~\ref{fig:density_profile} shows that the the number of Sgr members in the leading arm (closer to the Galactic plane) is $\sim10-15\%$ lower than in the trailing arm within the same distance from its centre. This indicates that our catalogue is somewhat incomplete at low Galactic latitudes, but it remains very pure: the sheer excess of field stars means that only very few actual Sgr members have a high enough likelihood ratio to be classified as such. This constrasts with a more traditional selection procedure based on fixed boxes in the CMD and PM spaces, which becomes more contaminated as the density of field stars increases. The entire catalogue of candidate members is available in the electronic form at \url{https://zenodo.org/record/3874830}.

\section{Line-of-sight kinematics}  \label{sec:vlos}

Some of the brightest Sgr members have line-of-sight velocity measurements from the \Gaia RVS instrument ($\sim 100$ member stars in the entire region). Much larger spectroscopic samples are available from other sources, of which we consider three complementary catalogues.  \citet{Penarrubia2011} observed $\sim 2000$ stars in 6 fields up to $4^\circ$ from the Sgr centre along the major axis and up to $2^\circ$ along the minor axis; almost 90\% of them are actual Sgr members. \citet{Frinchaboy2012} observed $\sim 2300$ stars in 24 fields across a large fraction of Sgr dSph: from $-4^\circ$ to $12^\circ$ along the major axis, $-3^\circ$ to $5^\circ$ along the minor axis, and a few diagonally situated fields. Roughly a half of these stars are Sgr members. Finally, several fields of the APOGEE spectroscopic survey are within the footprint of Sgr, and contain $\sim 900$ member stars, mostly in the central region; some fraction of this sample was analyzed in \citet{Majewski2013}. The CMD distribution of stars from these spectroscopic samples is shown on the left panel of Figure~\ref{fig:CMD}. There exist other spectroscopic datasets for the Sgr galaxy, but they are either of a lower precision \citep[e.g.,][]{Ibata1997} or cover only its nucleus \citep{Bellazzini2008,AlfaroCuello2020}, so we do not use them.

The spectroscopic datasets discussed above actually have a significant number of stars in common -- around 100 stars are found in all three spectroscopic samples (excluding \Gaia RVS stars, which are generally brighter and more scattered across the region). We find that the $v_\mathrm{los}$ measurements are largely consistent between independent datasets within error bars. The differences between velocities of individual stars are of order a few \kms, and the systematic offsets between entire samples are at a level of $1-2$~\kms, significantly smaller than the velocity dispersion of Sgr or the gradient of the mean $v_\mathrm{los}$ across the galaxy. We thus combine the information from all available sources, for a total of 3300 member stars with $v_\mathrm{los}$ measurements.

We group this spectroscopic sample into 36 Voronoi bins, with the majority (30) being located within $\sim 4^\circ$ from the Sgr centre, and remaining ones (mostly from the \citealt{Frinchaboy2012} sample) spread along the trailing arm up to $12^\circ$ from the centre. We compute the mean $v_\mathrm{los}$ and its dispersion in each bin, taking into account the measurement uncertainties. Since the values of $v_\mathrm{los}$ for individual stars were taken into account together with other astrometric and photometric properties when determining the membership probability, we do not need to impose further filters on the sample. The statistical uncertainties on the mean $v_\mathrm{los}$ are at the level $1-3$~\kms, depending on the number of stars in bins. We provide the derived values in Table~\ref{tab:vlos} and discuss the kinematic maps in Section~\ref{sec:kinematics}.

\section{Photometry, distance and 3d structure}  \label{sec:photometry}

\begin{figure*}
\includegraphics{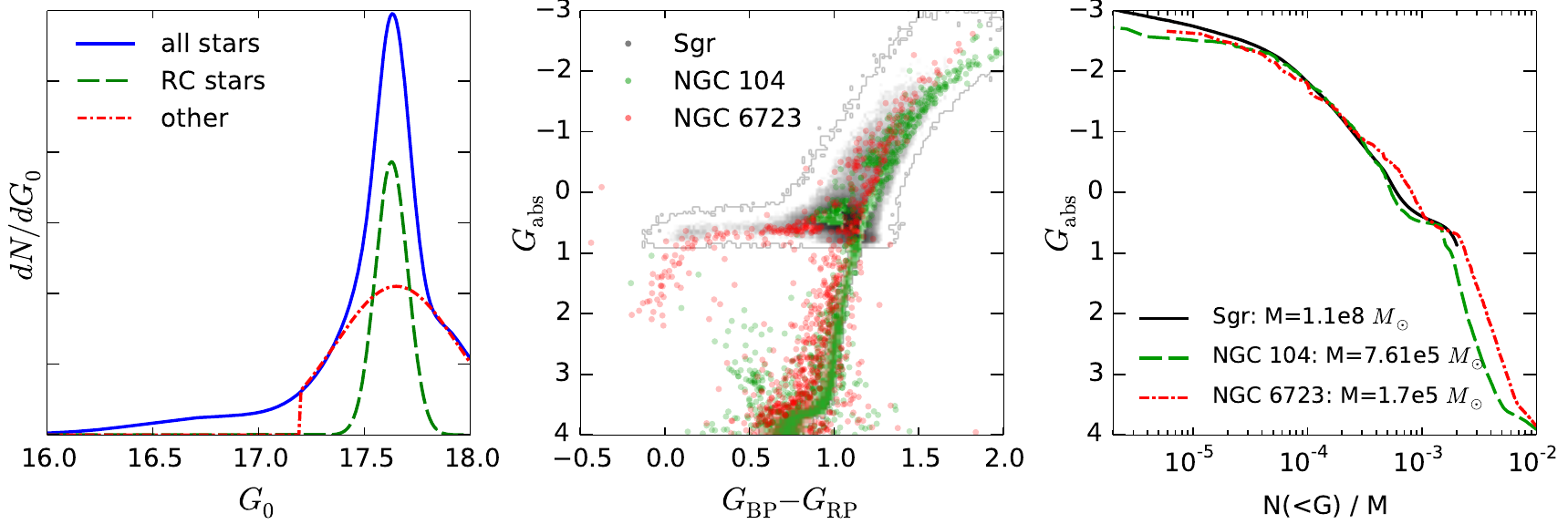}
\caption{\textit{Left panel:} distribution of extinction-corrected apparent magnitudes of $\sim25000$ Sgr stars in a circular region of radius $2^\circ$ located $4^\circ$ from the centre ($\alpha=288.5^\circ,\delta=-31.7^\circ$). Solid blue line shows the distribution of all stars, which is fit by two truncated Gaussians in the range $17.2<G_0<18$, shown by green dashed and red dot-dashed lines. We assume that the narrower Gaussian corresponds to RC stars, and use its mean and dispersion to estimate the photometric distance and thickness (this analysis is performed separately for each of the 200 Voronoi bins, containing $\sim1000$ stars per bin, and the results are shown in Panels H and J of Figure~\ref{fig:maps_data}).\protect\\
\textit{Centre panel:} CMD of Sgr (gray) compared to those of two globular clusters with comparable metallicity [Fe/H]${}\sim -0.9\pm 0.2$: 47 Tuc (NGC 104, green) and NGC 6723 (red). Magnitudes and colours are extinction-corrected and translated to the absolute magnitudes.\protect\\
\textit{Right panel:} cumulative number of stars brighter than the given absolute magnitude, normalized by the mass of the stellar system. The masses for the two globular clusters are determined by \citet{Baumgardt2019} from dynamical modelling, while the mass of the Sgr remnant is chosen to match the other two curves, and is estimated to be in the range $(1-1.2)\times10^8\,M_\odot$.
}  \label{fig:CMDfit}
\end{figure*}

It is apparent that the Sgr remnant is stretched along its orbit in projection, however, its 3d structure is less well known. \citet{Ibata1997}, based on the then-available photometric data, examined the magnitude distribution of Sgr stars around the red clump (RC), which serves as an absolute photometric reference point. They found no significant gradient of the RC magnitude (equivalently, variation of the distance) across the face of the galaxy, and adopted the distance $D=25\pm 2$~kpc. The dispersion of the Gaussian fit to the RC magnitude distribution was found to be $\sim 0.04$~mag, which corresponds to the upper limit on the thickness of $\sim 0.5$~kpc\footnote{In this section, we refer to the dispersion $\sigma$ of the Gaussian distribution as the thickness; the full width at half maximum is, as usual, 2.35$\sigma$.}, comparable to the width along the projected minor axis. Hence they concluded that the Sgr remnang has a strongly prolate cigar-like shape with axis ratios $\sim 3:1:1$, with its major axis roughly perpendicular to the line of sight.

Over the last two decades multiple groups employed a variety of stellar tracers to measure the structural properties of the Sgr remnant \citep[e.g.][]{Mateo1995,Mateo1998,Marconi1998}. Predictably, RR Lyrae have been used most extensively to measure the distance and  the total stellar mass of the dwarf and to estimate its metallicity and its spread \citep[see e.g.][]{Alcock1997,Cseresnjes2000, Cseresnjes2001, Kunder2009}. Most recently, \citet{Hamanowicz2016} used RR Lyrae stars from the OGLE catalogue to derive the distance and thickness of the Sgr core. The footprint of the catalogue covers only the central few degrees. They found a mean distance $D\simeq 26.7$~kpc with a systematic error of 1.3~kpc, and a (deconvolved) thickness (dispersion) along the line of sight $\sigma\sim 1$~kpc. \citet{Ferguson2020} used the same OGLE RR Lyrae catalogue, complemented with $\sim 800$ RR Lyrae stars from the \Gaia catalogue of variable stars, to study the 3d shape and orientation of the Sgr core, assuming it to be a triaxial ellipsoid with a Gaussian density profile. They concluded that the intrinsic shape of the Sgr core is indeed triaxial, with axis ratios $1:0.76:0.43$, with the major axis nearly perpendicular to the line of sight, and the intermediate axis parallel to it. Their estimate of the distance to the Sgr centre is slightly smaller, 26.4~kpc, and the length (dispersion of the Gaussian) of the major axis is 1.7~kpc. We adopt a distance $D_0=26.5$~kpc as our fiducial value.

We carried out the analysis of spatial variation of apparent magnitudes for RC stars, using the \Gaia photometry. For the RC stars, which constitute roughly half of our final sample, we determined the peak and width of their distribution as follows. We selected all stars with $17.2 < G_0 < 18$, $0.8 < G_\mathrm{BP,0}-G_\mathrm{RP,0}$ and membership probability above 0.8. Then for each of the 200 Voronoi bins, we construct a two-component truncated Gaussian mixture model for the distribution of stars in $G_0$, taking into account the sharp boundaries of the selection box (Figure~\ref{fig:CMDfit}, left panel). The dispersion of the narrower Gaussian component ranges from $\sim0.06$ in the centre to $\sim0.15$ in the trailing arm, corresponding to the thickness along the line of sight $\sim0.7-2$~kpc (Figure~\ref{fig:maps_data}, panel H). This is only an upper limit on the thickness, since the width of the Gaussian magnitude distribution is broadened by the intrinsic scatter in RC absolute magnitudes, which we did not attempt to subtract.

The mean magnitude is translated to the distance assuming an absolute G-band magnitude of RC stars of $0.47$ and plotted in Figure~\ref{fig:maps_data}, panel J. It is apparent that most of the trailing side of Sgr remnant is slightly more distant that its centre, but at distances $\gtrsim 15^\circ$ from the centre, the mean distance to stars starts to decrease. This apparently corresponds to the transition from the cigar-shaped remnant to the trailing arm of the stream, which has a minimum heliocentric distance $\sim 20$~kpc at $\sim 50^\circ$ from the centre.  We caution that the variation in apparent magnitudes is fairly small ($\lesssim 0.1$), and may be affected by the spatially varying reddening. Even though we did take it into account when computing extinction-corrected magnitudes and colours, there is still some variation in the mean colour of RC stars in the regions with high reddening, indicating possible biases in the photometric distance estimates. Nevertheless, as discussed in later sections, the features seen in the distance map and in the PM maps line up quite naturally, also indicating a transition between the remnant and the trailing arm around the same location.

The magnitude distribution of all Sgr stars can be used to infer its total luminosity and stellar mass, by comparing it with globular clusters of similar metallicity. Figure~\ref{fig:CMDfit}, centre panel, shows that the CMD of Sgr (at least its red giant branch including RC) is most similar to those of globular clusters with metallicity around $-0.9$; for comparison, we chose two clusters with large enough number of stars and low reddening -- NGC 104 (47 Tuc) and NGC 6723. These clusters have been extensively studied, and in particular, \citet{Baumgardt2019} fitted $N$-body dynamical models to the cluster kinematics and photometry, and estimated their total masses. 
We may infer the stellar (but not dynamical) mass of Sgr remnant by comparing the distribution of its stars in absolute magnitudes to these clusters, after normalizing the star counts by total masses. Figure~\ref{fig:CMDfit}, right panel, shows that a good match of the cumulative number of stars as a function of absolute magnitude is obtained for the stellar mass of Sgr around $10^8\,M_\odot$. We stress that this estimate does not assume that Sgr is in dynamical equilibrium, nor that its stellar mass is equal to the dynamical mass. These assumptions are only made for the globular clusters, and in addition we assume that the stellar mass functions of Sgr and the clusters are similar. Our approach neglects the fact that stars in Sgr are generally younger than in the globular clusters chosen for comparison.
We did not calculate the total luminosity of the Sgr remnant, but if its stellar mass-to-light ratio  is also similar to that of the clusters ($M/L_V \simeq 1.8$), then its V-band luminosity would be around $0.6\times10^8\,L_\odot$. For comparison, \citet{NiedersteOstholt2010} estimated it to be $(0.4\pm 0.06)\times10^8\,L_\odot$, and \citet{Majewski2003} quote a value twice as smaller. Their estimate is based on a different approach -- extrapolating the surface brightness profile after subtracting the MW foreground.

\section{Observed kinematics of the Sgr remnant}  \label{sec:kinematics}

\subsection{Choice of coordinates}

For the Sgr centre, we adopt the distance $D_0=26.5$~kpc, line-of-sight velocity $v_\mathrm{los,0}=142$~\kms, and PM $\mu_{\alpha,0}=-2.7$~\masyr, $\mu_{\delta,0}=-1.35$~\masyr.
We adopt the following values for the Solar position and velocity in the Galactocentric rest frame: $X=-8.1$~kpc, $V_{\{X,Y,Z\}}=\{12.9,\, 245.6,\, 7.8\}$~\kms \citep{Astropy}; the corresponding position and velocity of the Sgr centre are $\{X,Y,Z\}=\{17.5,\, 2.5,\, -6.5\}$~kpc, $V_{\{X,Y,Z\}}=\{237.9,\, -24.3,\, 209.0\}$~\kms.

Before presenting the kinematic maps, we need to define the coordinates and quantities to plot. The analysis described in Section~\ref{sec:membership} used the PM in the ICRS coordinates $\alpha,\delta$; however, the mean PM and their dispersion tensor can be equivalently transformed into any other celestial coordinates of choice $\chi,\xi$, using standard expressions for spherical geometry. It is natural to assign $\chi_0=\xi_0=0$ to the Sgr centre, but the orientation of the axes remains a free parameter. We now argue that it makes sense to align one of the coordinate axes (say, $\chi$) with the direction of the mean PM of the Sgr core on the sky plane, so that $\mu_{\chi,0} = \sqrt{\mu_{\alpha,0}^2 + \mu_{\delta,0}^2}$ and $\mu_{\xi,0}=0$. This direction is different from the major axis of Sgr (which itself very nearly coincides with its orbit on the sky plane), because the observed PM has a contribution from the solar motion. However, it is in these coordinates that the perspective effects have the most straightforward manifestation. 

Consider a situation when an isolated, non-rotating galaxy with isotropic velocity dispersion $\sigma$ is located at a distance $D_0$ and moves with a velocity $v_0$ relative to the observer, directed parallel to the $\chi$ axis. The mean PM of stars at a distance $D$ is $\mu_\chi=v_0/D$ and $\mu_\xi=0$, and its dispersion is $\sigma/D$ for both components. Since the galaxy has a finite thickness, one needs to integrate along the line of sight to obtain the PM dispersion $\sigma_\mu$. Assuming for definiteness that the density profile along the line of sight is a Gaussian with a dispersion $h \ll D$, it is easy to show that in the first approximation, the average PM is $\mu_{\chi,0}=v_0/D_0$, $\mu_{\xi,0}=0$, and its dispersion is
\begin{equation}  \label{eq:sigmaPM}
\sigma_\chi \;=\; \sqrt{\sigma^2 + (\mu_{\chi,0}\,h)^2}\, \big/ D_0,\quad
\sigma_\xi  \;=\; \sigma/D_0.
\end{equation}
In other words, the PM dispersion along the direction of motion is broadened by the spread in distances, while in the perpendicular direction it remains the same as if the galaxy had zero thickness. For Sgr, $v_0 \simeq 380$~\kms, $\sigma \simeq 13$~\kms, $h\simeq 1$~kpc, $D_0=26.5$~kpc, and both terms in the expression for $\sigma_\chi$ have comparable magnitudes. Therefore, the inflation of PM dispersion along the direction of motion due to perspective effects is very significant, and the alignment of the $\chi$ axis along the PM vector roughly diagonalizes the PM dispersion tensor, justifying our choice of the coordinate system.

Continuing with our toy example, if a galaxy moving as a solid body (with a spatially uniform mean velocity $\boldsymbol v_0$) subtends a finite region on the sky, the observed PM and $v_\mathrm{los}$ field will not be constant due to perspective distortions. Let $v_\mathrm{los,0}$ be the component of velocity along the line of sight passing through the galaxy centre ($\chi=\xi=0$), and $v_\mathrm{tan,0} \equiv \mu_{\chi,0}\,D_0$ be the velocity component parallel to the $\chi$ axis (the third component is $\mu_{\xi,0}=0$ by construction). Consider a star at a distance $D=D_0\,(1+\zeta)$, with the dimensionless parameter $\zeta\ll 1$ quantifying the offset from the mean distance to the galaxy, and sky-plane coordinates $\chi,\xi \ll 1$, moving with a total velocity $\boldsymbol v_0 + \boldsymbol u$. Its relative velocity with respect to the galaxy centre $\boldsymbol u$ has components $u_\chi,u_\xi,u_\zeta$ in three perpendicular directions. To a first order in $\chi,\xi$ and $\zeta$, the observed PM and $v_\mathrm{los}$ of this star are
\begin{equation}
\begin{array}{lll}
\mu_\chi &\!\!\!=\!\!\!&
\mu_{\chi,0}  - v_\mathrm{los,0}/D_0\,\chi - \mu_{\chi,0}\,\zeta + u_\chi/D ,\\[1mm]
\mu_\xi  &\!\!\!=\!\!\!&
\mu_{\xi,0}\, - v_\mathrm{los,0}/D_0\,\xi  - \mu_{\xi,0}\:\zeta\,+ u_\xi/D  ,\\[1mm]
v_\mathrm{los} &\!\!\!=\!\!\!&
v_\mathrm{los,0} + \mu_{\chi,0}\,D_0\,\chi + \mu_{\xi,0}\,D_0\,\xi + u_\zeta .
\end{array}
\end{equation}

The terms proportional to $\chi,\xi$ are caused by perspective effects: in the direction different from the galaxy centre, the line-of-sight velocity has a contribution from the centre-of-mass PM, and vice versa. The magnitude of these corrections is quite significant and can be larger than the relative velocity components, but it involves known quantities, and can be subtracted to obtain the "perspective-corrected" PM field:
\begin{equation}  \label{eq:PMcorr}
\begin{array}{lllll}
\mu_\chi' &\!\!\!\equiv\!\!\!& \mu_\chi + v_\mathrm{los,0}/D_0\,\chi  &\!\!\!=\!\!\!&
\mu_{\chi,0} - \mu_{\chi,0}\,\zeta + u_\chi/D ,\\[1mm]
\mu_\xi'  &\!\!\!\equiv\!\!\!& \mu_\xi  + v_\mathrm{los,0}/D_0\:\xi   &\!\!\!=\!\!\!&
\mu_{\xi,0}\, - \mu_{\xi,0}\;\zeta  + u_\xi/D .
\end{array}
\end{equation}
However, the term proportional to $\zeta$ also has an equally significant contribution, and cannot be corrected since the distance offset $\zeta$ is unknown. With our choice of $\mu_{\xi,0}=0$, this term is zero in the second row, so $\mu_\xi'$ does contain only the actual relative velocity field (still scaled by the unknown distance).

The line-of-sight velocity can also be corrected for the perspective effects, and does not contain any terms proportional to the unknown distance offset $\zeta$. Obviosly, this is possible only if the tangential component of the total velocity of the object's centre of mass relative to the observer $v_\mathrm{tan,0} = \mu_{\chi,0}\,D_0$ is known. When the object's PM is unavailable, people often use a ``partially corrected'' quantity -- so-called Galactic Standard of Rest (GSR) velocity $v_\mathrm{los,GSR}$. It is defined as the velocity that would be measured by an observer residing at the Solar position, but with zero velocity within the MW:
\begin{equation}  \label{eq:VlosGSR}
v_\mathrm{los,GSR} \equiv v_\mathrm{los} + \boldsymbol{v}_\odot\cdot \boldsymbol{n},
\end{equation}
where $\boldsymbol{v}_\odot$ is the 3d solar velocity in the MW rest frame, and $\boldsymbol{n}(\chi,\xi)$ is the unit vector in the direction of the given point $\{\chi,\xi\}$ on the celestial sphere. In other words, this correction involves only the solar velocity, but not the total centre-of-mass velocity of the object, and hence does not get rid of all perspective effects. By definition, $v_\mathrm{los,GSR}$ measures the velocity component \textit{in the direction of the observer}, and this direction varies across the object. Thus, even if all stars in a galaxy were moving with the same 3d velocity, the observed $v_\mathrm{los,GSR}$ would measure projections of this velocity onto different lines of sight, and hence would not be spatially uniform. Conversely, a constant $v_\mathrm{los,GSR}$ field does not imply the absense of internal rotation. Therefore, $v_\mathrm{los,GSR}$ does not have any advantages compared to $v_\mathrm{los}$ for describing the internal kinematics (in fact, a possible disadvantage is that the transformation between $v_\mathrm{los}$ and $v_\mathrm{los,GSR}$ obviously depends on the adopted spatial velocity of the Sun, which may differ between studies). However, it turns out that the mean value of $v_\mathrm{los,GSR}$ varies only mildly across the Sgr remnant, unlike either the heliocentric $v_\mathrm{los}$ or the internal velocity $u_\zeta$, which is nothing more than a fortuitous coincidence. For this only reason, we will be plotting $v_\mathrm{los,GSR}$ to highlight the small differences between observations and models, which otherwise would have been swamped by the strong gradients caused by perspective effects.

\subsection{Analysis of kinematic maps}

\begin{figure*}
\includegraphics{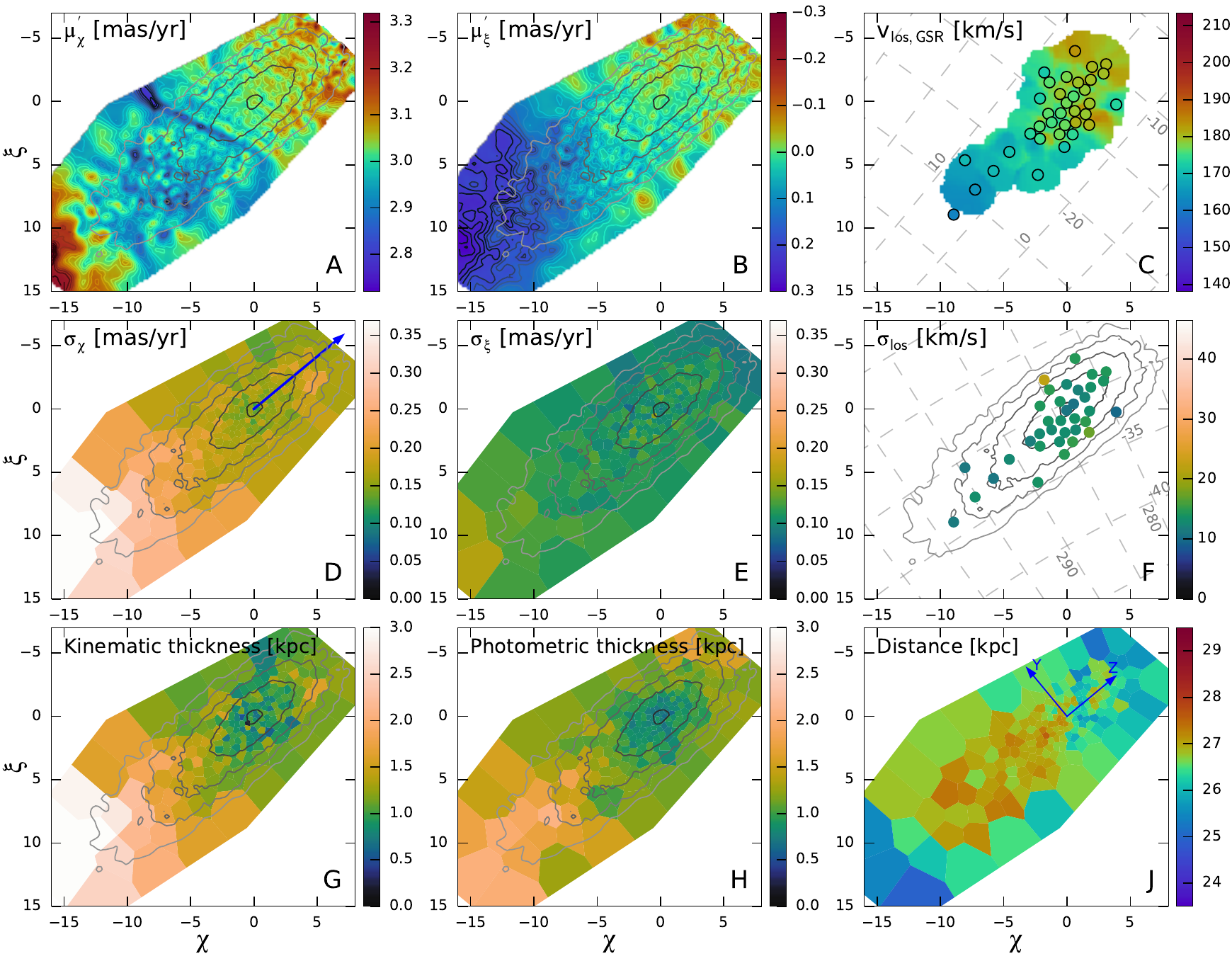}
\caption{Kinematic maps of the Sgr remnant. Coordinates are aligned with the apparent (non-reflex-corrected) motion of the object on the sky (the motion is in the direction of increasing $\chi$); the true velocity of the Sgr centre (blue arrow in panel D) points roughly towards the Galactic plane, which lies slightly beyond the top right corner. A grid of Galactic coordinates is shown in Panel C, and equatorial coordinates -- in panel F. Blue arrows in Panel J show two axes of the coordinate system aligned with the orbital plane, which is used in Figure~\ref{fig:maps_internal}; the third axis (X) points away from the eye. Most panels also display the surface density, with contours logarithmically spaced by 0.4 dex (i.e., one magnitude).\protect\\
Panels A and B show the perspective-corrected mean PM $\mu_\chi'$, $\mu_\xi'$ (Equation~\ref{eq:PMcorr}); panels D and E -- the PM dispersions $\sigma_\chi$, $\sigma_\xi$. The first of these values is inflated due to a non-negligible thickness of the galaxy, and the difference between the two dispersions (panel G) can be used to estimate the ``kinematic thickness'' $h_\mathrm{kin}$ (Equation~\ref{eq:hkin}). It increases from $\sim 1$~kpc in the centre to $\sim 3$~kpc towards the trailing arm, and closely resembles the photometrically estimated thickness (panel H). Panels C and F show the Galactic Standard of Rest line-of-sight velocity $v_\mathrm{los,GSR}$ (Equation~\ref{eq:VlosGSR}) and its dispersion. Panel J shows the photometric distance estimate, which correlates with the features seen in the mean PM $\mu_\chi'$ parallel to the direction of motion due to perspective effects.
Colour scales for $v_\mathrm{los,GSR}$ and $\sigma_\mathrm{los}$ match those of PM for an assumed distance $D_0=26.5$~kpc. 
}  \label{fig:maps_data}
\end{figure*}

Figure~\ref{fig:maps_data} summarizes all observational data on the kinematics of the Sgr core. Top row shows the mean perspective-corrected PM $\mu_\chi'$, $\mu_\xi'$ (panels A, B) and $v_\mathrm{los,GSR}$ (panel C). As discussed above, $\mu_\xi'$ contains only the internal velocity component $u_\xi/D$; it is reasonably flat over the main body of the dwarf and shows a steady gradient with $\chi$ at the edges of the remnant where the stars move away from the centre into leading and trailing arms. The other component, $\mu_\chi'$, has contributions both from the internal velocity component $u_\chi/D$ and from the distance gradient $-\mu_{\chi,0}\,\zeta$. The sharp increase in $\mu_\chi'$ happening in the trailing arm $\sim 15^\circ$ from the Sgr centre indicates the sudden drop in the mean distance to stars, not the change in their internal velocity. This is corroborated by a similar feature in the photometric distance map (panel J). Physically, this corresponds to the transition between the Sgr remnant itself, which is tilted with respect to the line of sight and its own orbit, and the unbound tail, which is roughly parallel to the orbit, but lies at a larger distance. The heliocentric distance to the Sgr orbit decreases towards the trailing arm, but the mean distance to stars in the remnant increases until the transition zone.

Artifacts from the \Gaia scanning law manifest themselves as systematically offset mean PM in spatial regions $\sim 0.5^\circ$ across, most notably as a blue scar in the middle of the top left panel. The magnitude of these systematic errors is $\lesssim 0.05$~\masyr, and they do not obscure the real features seen in the data.

The middle row displays the PM and $v_\mathrm{los}$ dispersions. Panels D and E show the two components of PM dispersion tensor in these coordinates, confirming our expectations discussed above. The perpendicular component $\sigma_\xi$ reflects the true internal velocity dispersion and is nearly constant ($0.10-0.14$~\masyr) across the field of view, while the PM dispersion parallel to the direction of motion $\sigma_\chi$ is broadened by the distance spread (Equation~\ref{eq:sigmaPM}) and ranges from 0.15 to more than 0.35~\masyr. We may use the above toy example to estimate the ``kinematic thickness'' of the Sgr remnant. Assuming that the velocity dispersion is isotropic (which, as we shall see, is not quite true), the thickness is given by the difference between the two components of PM dispersion:
\begin{equation}  \label{eq:hkin}
h_\mathrm{kin} \equiv D_0\, \sqrt{\sigma_\chi^2 - \sigma_\xi^2} \,\big/ \mu_{\chi,0}.
\end{equation}
This quantity is plotted on the panel G, and resembles qualitatively the photometric thickness map (panel H), with lower values $\sim 1$~kpc in the centre and rapidly increasing towards the trailing arm. We also checked that the off-diagonal component of the PM dispersion tensor in these coordinates is indeed small.

The Galactic Standard of Rest line-of-sight velocity (panel C) is only measured in a small number of spatial bins, plotted as dots; to guide the eye, we also show a continuous $v_\mathrm{los,GSR}$ map obtained by interpolating among nearest 100 stars at each location. There is a mild gradient of $v_\mathrm{los,GSR}$ parallel to the major axis of Sgr, but as discussed above, this quantity does not have a straightforward physical interpretation by itself. Its dispersion, however, is a real physically relevant quantity, and is remarkably constant across the galaxy (panel F). $\sigma_\mathrm{los}$ lies in the range $12-14$~\kms, and is slightly lower in the centre \citep{Majewski2013}. Remarkably, the PM dispersion $\sigma_\xi$ translated into \kms is very similar to $\sigma_\mathrm{los}$. However, one cannot conclusively interpret this as a sign of isotropy, since the third velocity dispersion component contributes only a fraction of the PM dispersion $\sigma_\chi$, and hence cannot be measured directly.

However informative these kinematic maps are, they are still not sufficient to reconstruct the internal velocity field $\boldsymbol u$ within the galaxy. Two of its components, $u_\xi$ and $u_\zeta$, can be read off the panels B and C ($\mu_\xi'$ and $v_\mathrm{los,GSR}$); however, the PM component $\mu_\chi'$ contains entangled information about both the component of velocity $u_\chi$ parallel to the apparent direction of motion and the mean distance to stars, which is not known to a sufficient accuracy to be subtracted. Therefore, a proper dynamical model is needed to interpret the observations.

\section{$N$-body models}  \label{sec:models}

\subsection{General considerations}

It is clear that the Sgr remnant is a heavily perturbed stellar system, and modelling it within the steady-state approximation would be inadequate. Instead, we explore evolutionary models of a disrupting satellite around the Galaxy, constraining them to have the present-day position and velocity of the Sgr remnant. 

We set up an equilibrium model for the Sgr galaxy, as described below, using the \textsc{Agama} framework \citep{Vasiliev2019a}. We then place it roughly in the apocentre of its orbit $\sim 2.5$~Gyr ago, so that it completes three pericentre passages before arriving at its present position (shortly after the third passage). This time interval allows to impose enough tidal perturbation to the remnant, while taking into account various practical considerations (e.g., the longer the simulation time, the more difficult it becomes to aim precisely at the given final state, given the dramatic mass loss). Of course, a real Sgr would have started its evolution some 10~Gyr ago from a much larger initial radius and having a much larger mass (up to $10^{11}\,M_\odot$, according to \citealt{Jiang2000} or \citealt{Gibbons2017}) than our adopted range of initial mass at the time 2~Gyr ago ($\sim 10^9\,M_\odot$), but the entire evolution is outside the scope of the present study; we are only concerned with the present-day state of the Sgr remnant.

We evolve the Sgr galaxy under its own self-gravity plus the static external tidal field of the Galaxy, assuming that the latter is fixed and not perturbed. We ignore the effect of dynamical friction and the response of the MW to the gravitational tug from its largest satellite -- Large Magellanic Cloud (LMC). For our range of initial masses, dynamical friction would change the orbit parameters by less than 10\% per orbital period. Likewise, the reflex motion and the distortion of the MW halo introduced by LMC \citep[see e.g.][]{Garavito2019,Erkal2019} are important for modelling the Sgr stream \citep[e.g.][]{VeraCiro2013,Gomez2015}, but this is not the goal of the present study. Instead, we adopt a reasonably realistic MW model, with parameters drawn from an ensemble of Monte Carlo samples from \citet{McMillan2017}, but with a less massive spherical halo than in their best-fit model; the circular velocity ranges from 225 to 185~\kms between 15 and 60~kpc. In this potential, the trailing arm of the Sgr stream aligns well with the observations, but the leading arm plunges back into the Galactic disc too early; it is known that a single spherical potential cannot simultaneously fit both arms of the stream \citep[see e.g.][]{Helmi2004,Johnston2005,Law2010}. However, it provides a good fit to the Sgr remnant.

We run the simulations with the $N$-body code \textsc{gyrfalcON} \citep{Dehnen2000}, which is included in the \textsc{Nemo} framework \citep{Teuben1995}, with the external potential of the MW provided by the \textsc{Agama} plugin for \textsc{Nemo}. The number of particles is a few$\times10^5$, softening length is $\epsilon=0.05$~kpc (Plummer equivalent is 0.035~kpc), and the maximum timestep is $\sim 2$~Myr with further two levels of subdivision based on particle accelerations.

\subsection{Initial conditions for the Sgr galaxy}

\begin{figure*}
\includegraphics{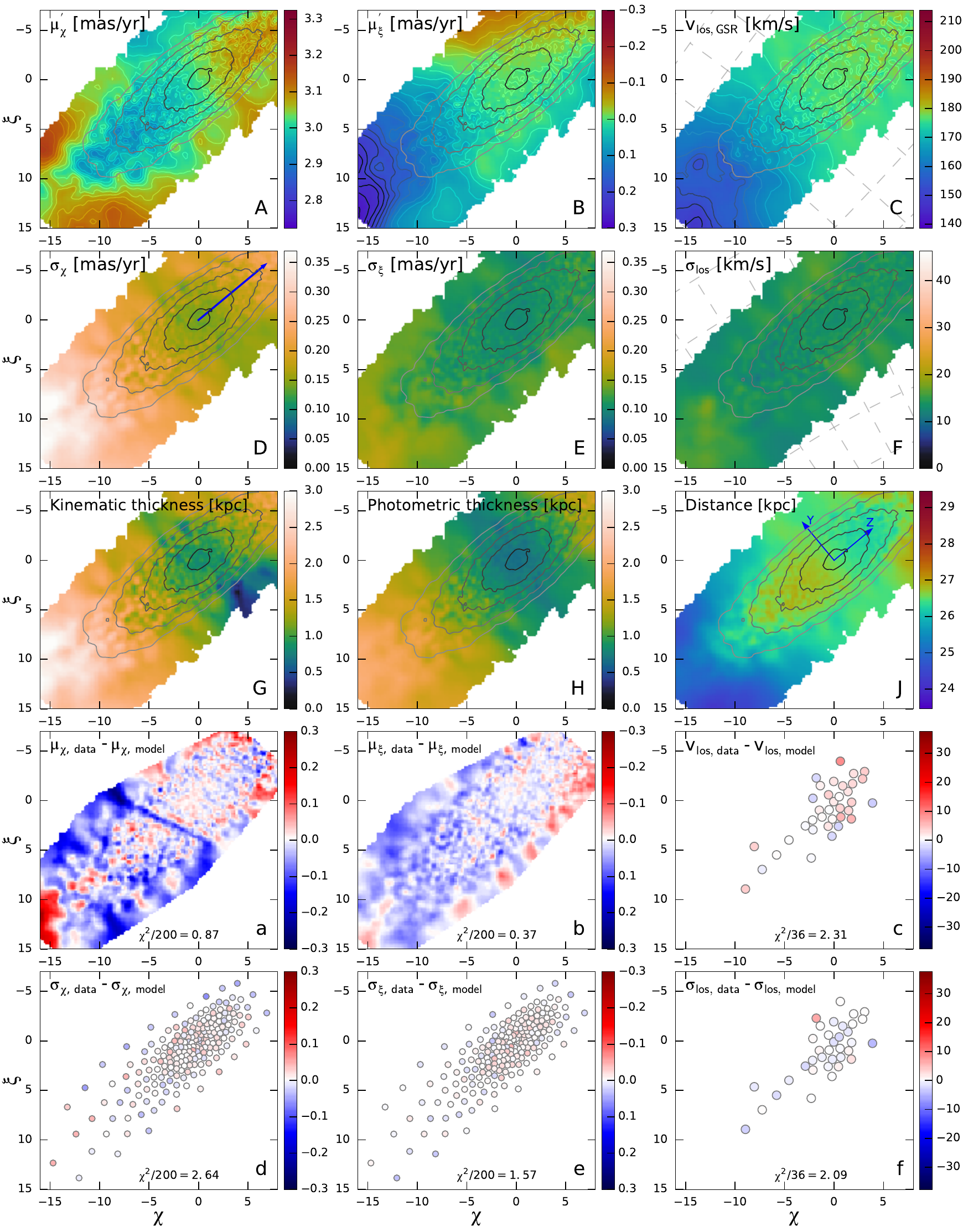}
\caption{Kinematic maps for one of the more successful models. The first three rows display the same quantities as the actual observations shown on Figure~\ref{fig:maps_data}, while the last two rows are the residuals for the first two rows.
\vspace*{-5pt}}  \label{fig:maps_model}
\end{figure*}

We explore many variants for the initial structure of the Sgr progenitor. Single-component models were found to be unable to match all observational constraints \citep[see][]{NiedersteOstholt2012,Gibbons2017}, so we concentrate on two-component models with more centrally concentrated stellar distribution embedded in a somewhat more extended dark halo. We consider both spherical and non-spherical stellar profiles, and the dwarf's halo was kept (nearly) spherical in all cases. The models vary in the degree of flattening, balance between rotation and dispersion, relative contribution of stars and dark matter to the total mass, and density profiles of the dark halo (cored or cuspy). The spherically-averaged stellar density is roughly the same in all of them and follows approximately an exponential or a King profile. This specific choice of the functional form has little impact on the density profile of the remnant after a couple of pericentre passages, when most of the mass has been stripped and the remaining one redistributed in response to tidal torques.

Spherical isotropic models are constructed using the Eddington inversion formula, while their flattened analogues are created with the iterative self-consistent modelling approach described in Section~5 of \citet{Vasiliev2019a}. The models are defined by distribution functions (DF) in action space, and the total potential corresponding to the density generated by the DF is computed iteratively. The properties of the model are thus determined by the choice of the DF family and its parameters. There are DF families suitable for disky stellar systems, with roughly exponentially declining surface density profiles and nearly constant thickness, or for more dispersion-supported oblate axisymmetric systems with rather flexible radial density profiles, which may also have net rotation.

Since we only follow the last stages of the Sgr disruption, its density profile cannot extend much beyond the tidal radius $r_\mathrm{tidal}$. For an orbit with an apocentre radius around 60 kpc, the mean density within $r_\mathrm{tidal}$ is approximately $10^6\,M_\odot\,\mathrm{kpc}^{-3}$, and at the pericentre radius of $\sim 16$~kpc this increases to $3.5\times10^7\,M_\odot\,\mathrm{kpc}^{-3}$. For realistic models, the initial tidal radius is $\sim10$~kpc; it drops to $\sim2$~kpc at the first pericentre passage, and to zero at the last (third) pericentre passage (i.e., even the central density of the remnant does not exceed the tidal limit). Hence there is no point of constructing initial models that extend significantly beyond the initial pericentric tidal radius, hence we put a Gaussian cutoff for the halo profile at a radius of a few kpc. For most of our models, the circular velocity curve peaks around 3~kpc with a velocity $\sim 50-60$~\kms, and the total mass lies in the range $(1-3)\times 10^9\,M_\odot$.

The nuclear star cluster M~54 has a mass $M_{54} \simeq 1.5\times10^6\,M_\odot$ \citep{Baumgardt2019}, dominating the gravitational field within a radius $G\,M_{54}/\sigma^2\sim 0.05$~kpc -- much smaller than the galaxy radius. We therefore ignore it in our models, although we did run one control simulation with a single massive particle added to the centre of the Sgr progenitor, finding no difference in the overall evolution.

\subsection{Fitting and analysing the simulations}

\begin{figure*}
\includegraphics{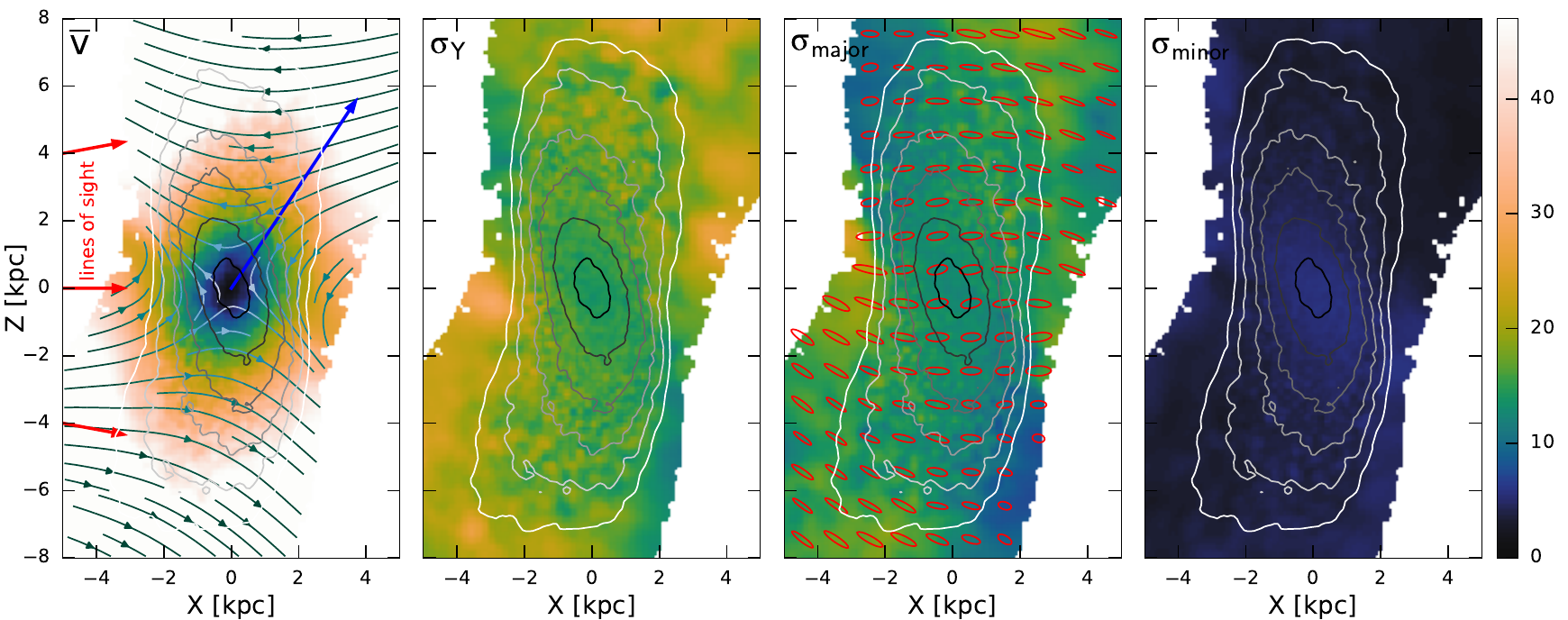}
\caption{Internal kinematics of the Sgr remnant (shown is same model as in Figure~\ref{fig:maps_model}, but there is little difference between models that fit the observations similarly well). Coordinates are aligned with the orbital plane of the Sgr galaxy, so that its angular momentum is antiparallel to the Y axis (which points away from the eye), and X axis is aligned with the line of sight (the observer is at $X\simeq -26.5$~kpc); the Y and Z basis vectors are shown as blue arrows in Panel~J of Figures~\ref{fig:maps_data} and \ref{fig:maps_model}. This coordinate system is only slightly rotated with respect to the Galactocentric coordinates. The orbit crosses the image plane diagonally from bottom left to top right; the velocity vector is shown by a blue arrow. Projected density contours are shown in all panels and are spaced logarithmically by 0.4 dex. Spatial and colour scales match those on the previous plots.\protect\\
\textit{Left panel:} mean relative velocity of stars with respect to the centre of the Sgr remnant. Colour indicates the magnitude of the velocity in \kms, while streamlines show its direction. It is clear that essentially no part of the remnant is moving as a solid body; not only it tumbles counter-clockwise (in the same sense as the orbital angular momentum), but is also strongly sheared, as evidenced by the X-shaped flow lines. \protect\\
\textit{Remaining panels} show the components of the velocity dispersion tensor indicated by colour. Centre-left panel displays the direction perpendicular to the orbit plane, and the other two panels -- the two eigenvalues (major and minor) of the in-plane velocity dispersion tensor, with its orientation also shown by red ellipses in the centre-right panel.
}  \label{fig:maps_internal}
\end{figure*}

The fitting strategy involves several steps. Each choice of the initial model still leaves room for rescaling its mass and radius. We pick up several values for the mass normalization, and for each mass determine the length scaling factor that produce a final result resembling the actual Sgr remnant, by running a grid of reduced-resolution simulations. At this stage, the most important comparison criteria are the remnant mass (stellar and dark), PM and $v_\mathrm{los}$ dispersions, and to some extent the shape; these are all linked and less sensitive to the accuracy of the final phase-space coordinates. Then we iteratively adjust the orbital initial conditions, as described in the Appendix~\ref{sec:orbit_ic}, running full-resolution simulations to match the present-day position and velocity; a single run takes up to half an hour for $N=5\times10^5$ particles. We also make small adjustments to the length scale of the model, to improve the fit for the velocity dispersions. Unlike other studies that fitted $N$-body models to MW satellites (Carina in \citealt{Ural2015} or Sculptor in \citealt{Iorio2019}), we have a larger amount of heterogeneous observational constraints and a more dramatic tidal evolution, so we assess the models only qualitatively, and the process involves a lot of subjective ``holistic'' judgement and manual labour. In total, we considered more than a hundred of models, of which only a small fraction were able to satisfy all available constraints even approximately.

The PM component perpendicular to the apparent motion, $\mu_\xi'$, is insensitive to the perspective effects, and all models produce very similar $\mu_\xi'$ maps, which also match observations rather well. On the other hand, the parallel component $\mu_\chi'$ is very sensitive to the distance gradient, hence providing strong constraints on the orientation and extent of the elongated remnant.  The angle between its major axis and the orbit needs to be around $45^\circ$ over a distance $\sim 5$~kpc to create the distinct dip in the PM map (panel A in Figure~\ref{fig:maps_data}) and a corresponding region of larger distances (panel J). This turned out to be the most challenging aspect of the system, and no models were able to match it perfectly.

Figure~\ref{fig:maps_model} shows the kinematic maps for one of the more successful models plotted in the same way as the real observations (Figure~\ref{fig:maps_data}), and additionally their residuals (differences from observed values). This model, and a number of other similarly looking models, is able to match qualitatively the features seen in the mean PM and $v_\mathrm{los}$ maps, and fits the dispersions reasonably well. The region of lower $\mu_\chi'$ extending up to $\sim 15^\circ$ from the centre towards the trailing arm (Panel A) is reproduced by the model, although not across the entire minor axis. The end of this region corresponds to the transition to the unbound and un-twisted part of the stream, which indicates that the 3d geometry of the remnant is represented adequately. The other PM component $\mu_\xi'$ and the line-of-sight velocity are not contaminated by perspective effects, and the mild gradient in the residual maps indicates some deficiencies in the fit, which may be partly alleviated by considering kinematically more complicated initial conditions (this particular model was initially spherical and non-rotating), or by adjusting the MW potential. The formal statistical uncertainties on PM dispersions are fairly small ($\lesssim 0.01$~\masyr), leading to mediocre values of reduced $\chi^2$ around 2; same is true for $v_\mathrm{los}$ and $\sigma_\mathrm{los}$ with uncertainties $\sim1-2$~\kms. At the same time the uncertainties on the mean PM are driven by \Gaia systematics and are much larger ($\sim0.04-0.05$~\masyr), hence the reduced $\chi^2$ values are less than unity. However, we believe that the variation of fit quality of the mean PM among different models is essential for interpreting the results, even if it appears to be statistically less significant.

\begin{figure*}
\includegraphics{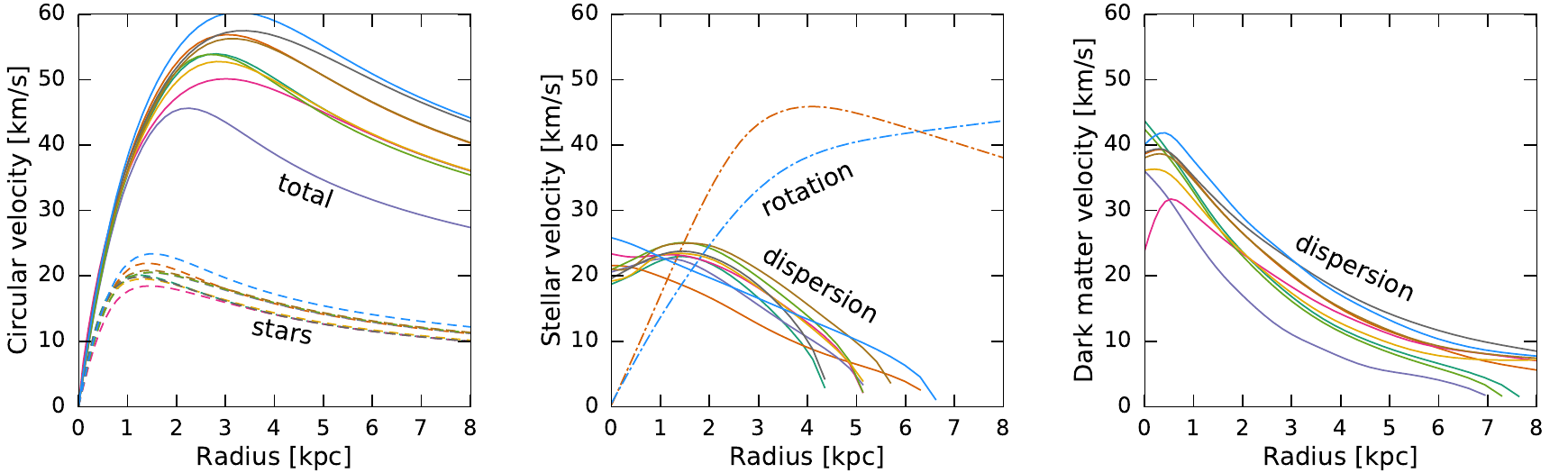}
\caption{Properties of initial models that fit the present-day Sgr remnant reasonably well after 2.5~Gyr of evolution (each model shown by a different colour).\protect\\
\textit{Left panel:} circular velocity curve $v_\mathrm{circ}(r) \equiv \sqrt{G\,M(<r) / r}$ corresponding to the total enclosed mass (solid) and stellar mass (dashed).\protect\\
\textit{Centre panel:} stellar velocity dispersion (solid) and mean rotational velocity (dot-dashed, only for the two rotating models).\protect\\
\textit{Right panel:} dark matter velocity dispersion. Note that since the dark matter is spatially more extended than stars, its velocity dispersion is significantly higher in the centre than the stellar dispersion, at least for cored dark matter profiles (all except the crimson-coloured curve, which plots a mildly cuspy model).
}  \label{fig:initprofiles}
\end{figure*}

A robust conclusion from the analysis of a large suite of models is that the Sgr core remained a bound stellar system until very recently. Models that were too large in size or not massive enough to withstand the tidal shocks at earlier pericenter passages produced a final configuration that was too stretched and more closely aligned with the orbit (i.e. without a distinct ``twist'' in the distance gradient). An example of such nearly disrupted system is given in Figure~\ref{fig:maps_model1}; the poor fit to $\mu_\chi'$ is evident. Conversely, if the initial configuration is too tightly bound and loses only relatively little mass, the final state is too spherical and either has too large velocity dispersion, or is too compact to match the observed extent of the remnant along its orbit (this is again most evident in the poor fit to $\mu_\chi'$). For instance, this was the case for the \citealt{Law2010} $N$-body model, which we also analyzed in the same way as our simulations. Figure~\ref{fig:maps_model2} shows the kinematic maps of a system similar to the \citealt{Law2010} model, but with slightly different orbital initial conditions tuned to match the present-day position and velocity of the Sgr remnant; their original model has very similar features. We also find that the dark halo of the Sgr remnant cannot be very cuspy: such models have either too high line-of-sight velocity dispersion or too small spatial extent. An example of such model is shown in Figure~\ref{fig:maps_model3}, having the velocity dispersion $\sim 3$~\kms higher than the actual galaxy. If we decrease the initial mass of this model to match the dispersion, it becomes too compact and the transition to the trailing arm occurs too early, producing a poor fit to $\mu_\chi'$.

Models with initially flattened and rotating stellar distribution may have several observable kinematic features. First, the gradient in the velocity components represented by $\mu_\xi'$ and $v_\mathrm{los,GSR}$ needs not be aligned with the major axis. However, there is little evidence for such a gradient in the observations (panels B and C in Figure~\ref{fig:maps_data}), putting an upper limit of a few \kms on the amount of rotation about the photometric major axis. Second, there may be a gradient in the distance along the minor axis, which would manifest itself in the $\mu_\chi'$ map; again, the data do not demonstrate such a gradient (panel A). Third, models in which the stellar distribution was rapidly rotating in the same sense as the orbital angular momentum develop a tidally induced bar upon passing the pericentre of their orbits \citep{Lokas2010}. These models can also match most of the features in the data (e.g., the orientation of the bar with respect to the orbital velocity), but they tend to have lower dispersions in both PM components, and moreover, they have a significant residual rotation about the photometric minor axis, manifested as a non-monotonic profile of $v_\mathrm{los,GSR}$ along the major axis (see, e.g., figure~14 in \citealt{Lokas2010}), in disagreement with the observations. An example of such model is given in Figure~\ref{fig:maps_model4}. In our experiments, models with an initial flattening greater than $1.5:1$ and maximum rotation velocity exceeding the central velocity dispersion were less successful in reproducing all observable properties. We conclude that models with significant initial flattening and rotation, although not strongly disfavoured, do not provide a noticeably better fit to observations than simpler spherical models.

Figure~\ref{fig:maps_internal} shows the intrinsic kinematic features of the Sgr remnant (they are very similar among the models that fit the observations reasonably well). One could immediately see that the remnant remains anything but an equilibrium configuration: the mean velocity of stars relative to its centre-of-mass steadily rises as one moves away from its centre, with a gradient $\sim 10$~\kms per kpc (left panel). Even though only a fraction of this mean velocity is directed radially, it is still clear that the system is expanding rapidly, and thus is far from a steady state. This implies that any analysis method based on the equilibrium assumption (such as Jeans equations) would give incorrect results regarding the mass distribution, and $N$-body simulations remain the only viable modelling approach. In addition, the streamlines of the mean velocity are not circles or ellipses, as would be in the case of a rotating system, but rather have a characteristic X shape indicating a shear (contraction in one direction and expansion in the perpendicular one). The velocity ellipsoid is very anisotropic, with the largest dispersion being in the direction perpendicular to the orbit plane (centre-left panel) closely followed by the dispersion roughly perpendicular to the orbital velocity (centre-right panel), and the remaining component being much smaller (right panel). This again defies expectations for a prolate stellar system, in which the dispersion along the major axis should be larger than in other directions -- and yet here it is exactly the opposite (note the orientations of the velocity ellipsoids in the third panel). We checked that this anisotropy remains in place even when we consider only bound particles. Unfortunately, it is nearly impossible to measure directly such a small dispersion in the direction parallel to the orbital motion, because, as discussed earlier, the observed PM dispersion is dominated by the spread in distances, not in the intrinsic velocity.

\begin{figure*}
\includegraphics{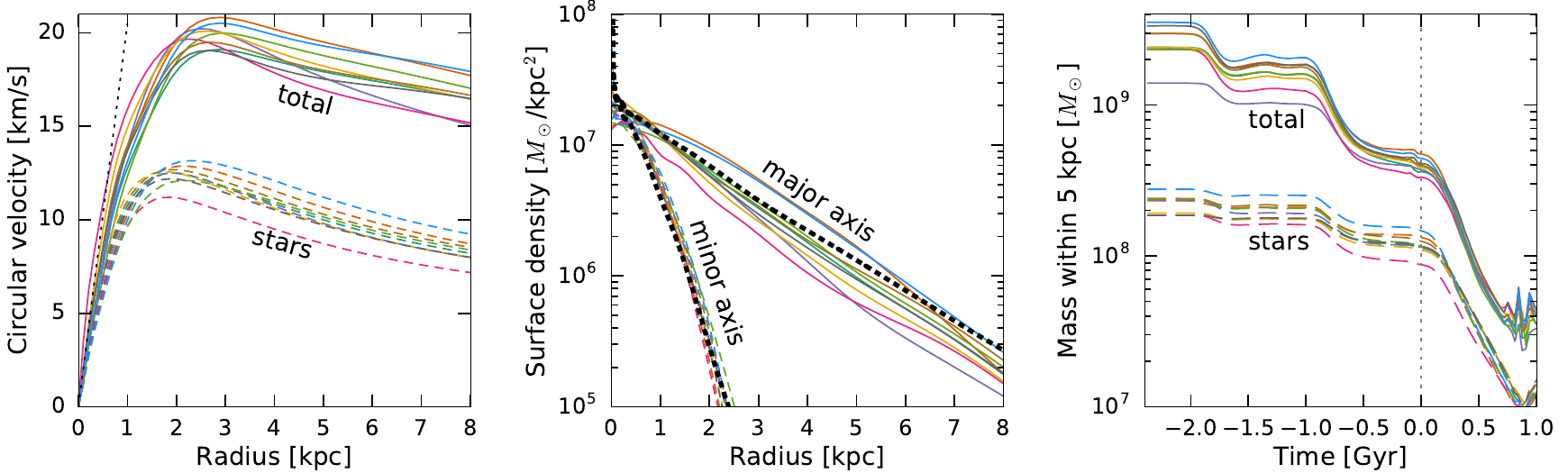}
\caption{Properties of some of the more successful models for the Sgr remnant (same set as in Figure~\ref{fig:initprofiles}, each model shown by a different colour).\protect\\
\textit{Left panel:} circular velocity curve $v_\mathrm{circ}(r) \equiv \sqrt{G\,M(<r) / r}$ corresponding to the total enclosed mass (solid lines) and stellar mass only (dashed lines). Black dotted line shows the tidal limit at the current position, indicating that the entire Sgr remnant is tidally disturbed (equivalently, its tidal radius is zero). \protect\\
\textit{Centre panel:} surface density profiles along the major (solid lines) and minor (dashed lines) axes. For comparison, the actual observations are plotted in black dots.\protect\\
\textit{Right panel:} time evolution of the total (solid) and stellar (dashed lines) mass enclosed within 5~kpc. The evolution is started some 2.5~Gyr before present at an apocentre; each successive pericentre passage leads to a tidal shock and causes a sudden mass loss, and the most recent passage actually initiates a complete unbinding of the remnant, which, nevetheless, will dissolve only gradually over the next Gyr.
}  \label{fig:profiles}
\end{figure*}

Figure~\ref{fig:initprofiles} illustrates the initial structural properties of a bunch of models with reasonably good fit quality. All of them have roughly similar initial density in the central parts (the rising part of the circular velocity curve), but are truncated at different radii and have different ratios of stellar to total mass. All but two models are spherical and non-rotating; the more strongly rotating one is actually a rather poor fit, but is included for the sake of diversity. All but one model have cored dark matter profiles (although the central dark matter density may exceed the stellar density). The only moderately cuspy model in this set is shown by a crimson line, and it still produces a noticeably worse fit than other models, exceeding the line-of-sight velocity dispersion constraints by $\sim 2$~\kms.
Notably, the stellar velocity dispersion in the centre is significantly lower than the dark matter velocity dispersion, and both decrease with time as the outer layers are progressively stripped away.

Figure~\ref{fig:profiles} shows the same set of models at the present moment. The stellar and total mass distribution is fairly similar between all these models, shown by the circular velocity curve (left panel), which measures the spherically-averaged enclosed mass profile. It peaks around $2.5-3$~kpc at a value $\lesssim 20$~\kms, and the stellar distribution is even more concentrated (stars dominate the total density within 1~kpc in most of these models). For reference, we also show the circular velocity corresponding to the mean density of $\sim 2.5\times10^7\,M_\odot\,\mathrm{kpc}^{-3}$, which is the tidal limit at the present location of the Sgr remnant. The fact that it lies just above all the models confirms the fact that the Sgr core is tidally disturbed down to its very centre (except the nuclear star cluster M~54, which we did not simulate). Centre panel shows surface density profiles along the major and minor axes, which match the observations fairly well. 

Since the transition between the remnant and the trailing arm of the stream occurs around $12-15^\circ$ from the centre, as indicated both photometrically and kinematically, we take the enclosed mass within a fiducial radius 5~kpc as our mass estimate. For most successful models, the total mass within this radius is $\sim (4\pm 1)\times10^8\,M_\odot$, of which the stars contribute around $10^8\,M_\odot$, in accordance with our photometric estimate (Section~\ref{sec:photometry}). The mass is mainly constrained by the velocity dispersion, but also produces configurations of an appropriate spatial extent and elongation.
We find that the total mass profile must be more extended than the stellar component, disfavouring mass-follows-light models. Only in this case the transition between the remnant and the stream occurs at large enough distances without producing an excessively high velocity dispersion in the centre. A similar argument leads to the preference of initially cored dark matter profiles, in which the stellar velocity dispersion in the centre is significantly lower than the dark matter dispersion, as shown in the centre and right panels of Figure~\ref{fig:initprofiles}. Stars need to be sufficiently ``cold'' to satisfy observational constraints, while the total mass has to be large enough for the galaxy to survive until present time and keep a sufficient spatial extent.

The inability of mass-follows-light models to match the properties of the remnant might also be alleviated in modified gravity theories such as MOND \citep{Milgrom1983}, since the internal accelerations in the Sgr remnant are $\lesssim10^{-9}\,\mathrm{cm/s}^2$ -- 10$\times$ lower than the fiducial value of the MOND acceleration constant $a_0$. However, one would need to take into account the external field effect of the Milky Way \citep[e.g.,][section 6.3]{Famaey2012}, since the Milky Way acceleration at the location of Sgr is comparable to $a_0$. Evolutionary simulations in MOND \citep[e.g.,][]{Thomas2017} would provide a stringent test of the theory in reproducing properties of both the Sgr stream and the remnant.

Right panel of Figure~\ref{fig:profiles} shows the time evolution of the enclosed mass within a fiducial radius 5~kpc (both stellar and total). Models started with rather different initial masses and concentrations all converge around the present-day mass of $\sim 4\times10^8\,M_\odot$ within this radius, of which stars contribute about a quarter. The ratio between stellar and total mass increases with time, and was $\lesssim 10\%$ at the beginning of simulation (models with a lower initial dark matter fraction were unable to maintain a sufficiently extended remnant without violating the velocity dispersion constraints). This trend contrasts with the findings of \citet{Penarrubia2008}, who simulated tidal stripping of dSph within cuspy dark haloes, and predict that the dark matter fraction in the remnant increases with time, until only a few stars remain in a very compact dark halo \citep[see also][]{Errani2020}. However, in the case of Sgr, we find that initially cuspy models are disfavoured by the data by the combination of velocity dispersion and spatial extent constraints.

\begin{figure*}
\includegraphics{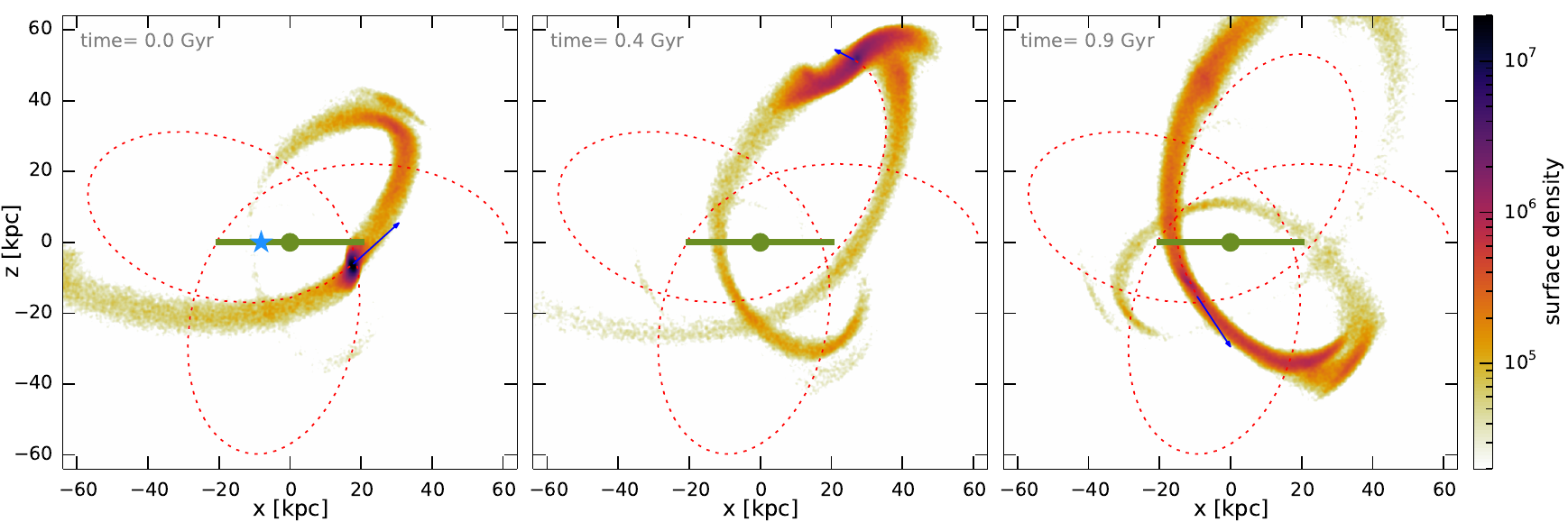}
\caption{Future evolution of the Sgr remnant. Shown is the surface density of stellar particles in the remnant and the stream in the Galactic side-on projection. The MW disc is perpendicular to the image plane (shown by a green line, with the Solar position marked by an asterisk), and the Sgr orbit lies close to the image plane. Left panel shows the present state, central -- next apocentre, and right -- next pericentre. The past trajectory of the Sgr centre-of-mass is shown by red dotted lines, and its velocity at each moment of time -- by a blue arrow. This plot shows the same fiducial model as in Figures~\ref{fig:maps_model}, \ref{fig:maps_internal}, but the evolution is very similar among all models: the Sgr galaxy becomes disrupted over the next orbital period.
}  \label{fig:evolution}
\end{figure*}

What is more interesting, though, is that when we continue the $N$-body simulation into the future, all models demonstrate very similar behaviour: the mass within a fixed radius drops precipitously -- in other words, the Sgr galaxy is completely disrupted over its next orbit (Figure~\ref{fig:evolution}). Although some concentration of mass close to its centre-of-mass is retained, it no longer remains a gravitationally bound system. Thus we conclude that we are witnessing the final demise of this once third-massive satellite of our Galaxy.
Its nuclear star cluster, M~54, will survive the disruption and join the group of ``peculiar'' Milky Way globular clusters with complex stellar populations, such as $\omega$~Cen \citep[e.g.,][]{Lee1999} or Terzan~5 \citep{Ferraro2009}, which are suspected to be remnants of larger stellar systems.

\subsection{Comparison with the literature}

Despite its proximity, the dynamical state of the Sgr dSph has been studied only in a few papers. The most widely known $N$-body model of \citet{Law2010} focused on reproducing the properties of the Sgr stream, not its core. As mentioned earlier, the remnant is much too compact in this model, and even though its line-of-sight velocity dispersion does not exceed the observational limits, the PM dispersion is too low, and the extent of the region of low $\mu_\chi'$ seen in the data is not reproduced by the model, since the transition from the nearly spherical core to the stream occurs too early (Panels A, G, J in Figure~\ref{fig:maps_model2}). They estimate the total mass of the remnant to be $2.5^{+1.3}_{-1.0}\times10^8\,M_\odot$ based on the velocity dispersion in the stream; the actual remnant mass in the $N$-body model is not quoted, but is likely smaller than our estimate ($\sim 4\times10^8\,M_\odot$) based on the lower velocity dispersion. The smooth monotonic trend of $v_\mathrm{los,GSR}$ along the major axis, seen in the data, is not quite reproduced by the model: it displays a ``kink'' (sudden steepening of the gradient) in the $v_\mathrm{los,GSR}$ within $2-3^\circ$ from the centre, where the remnant is intrinsically not rotating (Panel C in the above figure). Based on the analysis of our suite of models, this feature is characteristic of more concentrated systems that resist the tidal perturbation in their central parts.

\citet{Penarrubia2010} proposed a model in which the Sgr galaxy was initially a rapidly rotating disc embedded in a dark halo. This scenario could explain the bifurcation in the Sgr stream, but predicted a high degree of rotation in the remnant, which subsequently was not confirmed by newer observations in \citet{Penarrubia2011}. Notably, the model predicted a strong gradient in $v_\mathrm{los,GSR}$ along both axes (major and minor), whereas in reality the variation along the major axis has the opposite sign and is much shallower, as evidenced by larger-scale kinematic measurements of \citet{Frinchaboy2012}; the minor axis gradient in the data has the same sign but is also much weaker than in the model. We also analyzed the PM field predicted by this model and found it to be similarly discrepant with the observational data.

\citet{Lokas2010} presented another scenario with a disky Sgr progenitor, whose internal angular momentum was initially nearly co-aligned with its orbital angular momentum (i.e., the rotation is prograde with respect to the orbit). In this case, a strong bar perturbation is induced during a pericentre passage, and they find that after a second passage, their model provides a good match to the observations (in particular, the mean $v_\mathrm{los}$ and its dispersion measured by \citealt{Frinchaboy2012}). A more detailed analysis of their figures 10 and 11 suggests that the remnant is still too hot kinematically ($\sigma_\mathrm{los}\sim15-20$~\kms as opposed to the observed values $12-15$~\kms), and the $v_\mathrm{los,GSR}$ profile along the major axis (figure~14) is non-monotonic, unlike the observed one, although the disagreement is at the level of only a few \kms. The enclosed mass within 5~kpc is $\sim 4\times10^8\,M_\odot$, similar to our estimates, but it is more centrally concentrated (the peak circular velocity of $\sim 21$~\kms is attained at $\sim 1.6$~kpc in their model, as opposed to $2.5-3$~kpc in our models). Consequently, their model is far from being tidally disrupted, and survives for at least one more orbital period in a bound state. 

In the \citet{Lokas2010} scenario, the elongated shape of the remnant is caused by a tidally induced bar, which, as they argue, could only appear in a model with an initially prograde rotation, but not in an initially spherical system or in a retrogradely spinning disc. However, many of our models were initially non-spinning and spherical, and yet due to a strong tidal perturbation they acquire a bar-like shape with kinematics consistent with observations. We believe that this difference is caused by a lower concentration of our models, necessitated by a relatively low observed velocity dispersion. Coincidentally, this also introduces just a right amount of rotation in the remnant, reproducing the smooth trend of $v_\mathrm{los,GSR}$ along the major axis. While we do not rule out a possibility of an initially rotating Sgr progenitor, it appears to be unnecessary to explain the shape and kinematics of its remnant (although still might be needed to reproduce some features of the Sgr stream).

\subsection{Limitations of the model}

The $N$-body models considered in this paper were specifically designed to reproduce the properties of the Sgr remnant, not the stream, unlike most existing studies. Our treatment of the Sgr galaxy orbit is rather simplistic: all that we require is to arrive at the present location with the velocity consistent with observations. We do follow the Sgr galaxy as a live $N$-body system, however, the MW is represented as a static external potential, we ignore the effects of the LMC, and limit ourselves to one particular choice of the MW potential. We also neglect the dynamical friction, which is likely unimportant over the last 1 Gyr owing to a relatively small mass of the remnant, but certainly plays a role at earlier times.

We also start our simulations relatively recently ($2-2.5$ Gyr ago) and with an already truncated density profile of the Sgr progenitor. As a result, it loses relatively little mass (factor of $1.5-2$) after the first pericentre passage, but much more (factor of $2-5$) after the second (penultimate) one. Our models do not necessarily represent the mass loss history particularly well at early times, but they should be more reliable over the last orbital period and can be used to forecast the future evolution, being well constrained by the present-day state of the Sgr remnant. In short, the 2 Gyr of evolution is just a device to produce a realistically looking tidally perturbed model, which is then compared with observations. Our approach is also driven by practical considerations: we find it vitally important to obtain a very accurate fit to the present-day position and velocity in order to adequately compare the kinematics of different models. A longer simulation period and incorporation of additional perturbations on the orbit will make this task still harder.

The question remains whether the parameters responsible for the details of the preceding evolution (i.e., potential, orbit shape, mass loss history) can leave a discernible signature in the present-day state of the remnant. Based on our preliminary experiments, the answer is likely positive, and the models can and should be refined while adapting to other observational constraints on the structure of the Sgr stream. However, we believe that the main features of the Sgr remnant are unlikely to change qualitatively.

\section{Summary}  \label{sec:summary}

We presented a detailed analysis of the Sgr galaxy core, using the data from the \Gaia DR2 catalogue and other existing spectroscopic datasets. We developed a multidimensional mixture model to classify the catalogue into candidate Sgr members and field stars, which simultaneously produces astrometric kinematic maps of the Sgr galaxy. The final list of candidate members contains $\sim 2.6\times10^5$ stars with extinction-corrected $G$-band magnitude brighter than 18 (up to and including the red clump). The observational data summarized on Figure~\ref{fig:maps_data} already reveal a number of important features pertaining to both the spatial and kinematic properties of the Sgr remnant and its transition into the Sgr stream. A more detailed interpretation was conducted with the help of a large suite of $N$-body models for the Sgr remnant, which were evolved in the Galactic tidal field for 2.5 orbital periods preceding the most recent pericentre passage (which occurred $\sim30$~Myr ago). The results of this analysis can be summarized as follows.

\begin{itemize}
\item We estimated the total stellar mass of the Sgr remnant to be $\sim10^8\,M_\odot$ from its photometry.
\item At the same time, the total mass of the Sgr remnant within 5~kpc from its centre is a factor of four higher. We find that a mass-follows-light model is a worse match for the observational constraints compared to models with a more extended dark halo than the stellar distribution, and that the models with an initially cored dark matter profiles are preferred by the data.
\item The 3d shape and orientation of the Sgr remnant is strongly constrained by the imprint it leaves in the PM field. We find that the remnant is a prolate structure tilted at $\sim45^\circ$ with respect to its orbit, which transitions into the tidal stream beyond $\sim5$~kpc from its centre. This is also supported by photometric data, although we consider them to be less reliable.
\item The observed cigar-like shape is caused by the Galactic tidal field and is well reproduced even by models that start as spherically symmetric, without the need to invoke initial flattening or rotation.
\item The combination of a relatively low velocity dispersion with an extended prolate shape strongly suggests that the Sgr remnant ceased to be gravitationally bound after the most recent pericentre passage, and will gradually dissolve over the next orbital period.
\item The remnant is significantly out of equilibrium to render the classical dynamical modelling methods useless.
\end{itemize}

Our models were specifically tailored to reproduce the observed properties of the Sgr remnant, not the stream, and are a poor fit to the latter. It is likely that a successful reconstruction of the stream would require to add more physical ingredients influencing the Sgr orbit, which could also affect the properties of the remnant. Nevertheless, its global features are tightly constrained by abundant observational data, and should be taken into account by any study focusing on the Sgr stream or on the perturbations in the MW disc produced by the Sgr galaxy.
\vspace{5mm}

We thank Jorge Pe\~narrubia and Denis Erkal for providing snapshots of their simulations and for enlightening discussions, and the anonymous referee for a speedy and insightful report. 
We used the Whole Sky Database (wsdb) created and maintained by Sergei Koposov with financial support from the Science \& Technology Facilities Council (STFC).
This work uses the data from the European Space Agency mission \Gaia (\url{https://www.cosmos.esa.int/gaia}), processed by the \Gaia Data Processing and Analysis Consortium (\url{https://www.cosmos.esa.int/web/gaia/dpac/consortium}).
This work started at the Kavli Institute of Theoretical Physics during the programme ``Dynamical Models for Stars and Gas in Galaxies in the Gaia Era'', which was supported in part by the National Science Foundation under Grant No. NSF PHY-1748958.

\textbf{Data availability:} the entire catalogue of candidate Sgr members is available in the electronic form at \url{https://zenodo.org/record/3874830}, and the derived kinematic properties are provided in tables \ref{tab:pm}, \ref{tab:vlos} as supplementary online material.


\appendix
\section{Determination of the orbit initial conditions}  \label{sec:orbit_ic}

The $N$-body simulations discussed in Section~\ref{sec:models} start $\sim 2.5$ orbital periods ago, and are constrained to match the present-day centre-of-mass position and velocity of the Sgr remnant, using the procedure detailed below.

The first guess for the initial conditions (IC) of the Sgr orbit comes from integrating the orbit backwards in the static MW potential; however, the actual trajectory of a disrupting satellite deviates from a test-particle orbit, necessitating further refinement of the initial conditions. We employ the standard Gauss--Newton iterative procedure to find the orbital IC $\boldsymbol w_\mathrm{in} \equiv \{\boldsymbol x, \boldsymbol v\}_\mathrm{in}$ leading to the given final position and velocity $\boldsymbol w_\mathrm{end}$ after a fixed time $T_\mathrm{end}$. For a given choice of $\boldsymbol w_{\mathrm{in},0}$, we follow an ensemble of simulations with slightly offset IC $\boldsymbol w_{\mathrm{in},k}, k=1..K$, which produce the end states $\boldsymbol w_{\mathrm{end},k}$. Then the Jacobian matrix $\mathsf J \equiv \partial \boldsymbol w_\mathrm{end} / \partial \boldsymbol w_\mathrm{in}$ is approximated by finite differences: $\mathsf J \approx \delta\mathsf w_\mathrm{end}\, \delta\mathsf w_\mathrm{in}^{-1}$, where the columns of matrices $\delta\mathsf w_{\dots}$ contain the difference vectors $\boldsymbol w_{\dots,k}-\boldsymbol w_{\dots,0}$. Finally, the next choice of IC is given by $\boldsymbol w_\mathrm{in,0}^\mathrm{(new)} = \boldsymbol w_\mathrm{in,0} - \mathsf J^{-1}\,(\boldsymbol w_\mathrm{end,0} - \boldsymbol w_\mathrm{end}^\mathrm{true})$.

The procedure outlined above is a general way of solving nonlinear equation systems, but to make it practical in the present case, a few adaptations were made. Since the Sgr remnant has just passed the pericentre, its position and velocity are rapidly varying. The main effect of perturbing the IC is the slight change in the orbital energy and the corresponding change in flight time. After 2 Gyr of evolution, this translates to the final states of perturbed orbits being stretched along the trajectory, making the Jacobian extremely degenerate, with condition number exceeding $10^4$. On the other hand, this orbital motion is fairly predictable and may be treated separately from the perturbations perpendicular to the orbit.

We introduce two auxiliary coordinate systems for the initial and final states, aligned with the position and velocity vectors of the unperturbed orbit. Namely, $\boldsymbol w(t) = \boldsymbol w_{\mathrm{in},0} + \mathsf B_\mathrm{in}\,\boldsymbol p(t) = \boldsymbol w_\mathrm{end}^\mathrm{true} + \mathsf B_\mathrm{end}\,\boldsymbol q(t)$, where $\boldsymbol p(t)$ and $\boldsymbol q(t)$ are the phase-space coordinates of an orbit in either of the two auxiliary systems, and the orthogonal matrices $\mathsf B_\mathrm{in}$, $\mathsf B_\mathrm{end}$ are defined in such a way that the orbital motion occurs along the first component of $\boldsymbol p$ or $\boldsymbol q$ at $t=0$ or $t=T_\mathrm{end}$, respectively. The first column of $\mathsf B_\mathrm{in}$ is the unit-normalized\footnote{Since the position and velocity have different units, we introduce dimensional scaling factors before rotating the 6d phase space, choosing them to approximately match the magnitude of position and velocity variations -- in this case, using 1 kpc and 10 \kms as scale factors.} time derivative of the 6d phase-space coordinates, i.e., $\{\boldsymbol v_{\mathrm{in},0}, -\partial\Phi/\partial\boldsymbol x|_{\boldsymbol x = \boldsymbol x_{\mathrm{in},0}}\}$, and the remaining columns are all orthogonal to the first column and between themselves, but otherwise arbitrary. Similarly, $\mathsf B_\mathrm{end}$ is defined by the velocity and acceleration at $\boldsymbol w_\mathrm{end}^\mathrm{true}$. 

The IC of the baseline orbit is thus $\boldsymbol p_{\mathrm{in},0} \equiv \boldsymbol p_0(t=0) = \boldsymbol 0$, and the ICs of $K$ perturbed orbits $\boldsymbol p_{\mathrm{in},k},\; k=1..K$ are confined to the 5d subspace defined by setting the first component $p_k^{(1)}$ of each vector to zero, i.e., orthogonal to the unperturbed orbit at the initial point.
The final states $\boldsymbol q_{\mathrm{end},k} \equiv \boldsymbol q_k(T_\mathrm{end})$ of all $K+1$ orbits (including the unperturbed one) do not necessarily have zero in their first component (of course, the goal is to have $\boldsymbol q_{\mathrm{end},0} = \boldsymbol 0$, but we are searching for this solution iteratively). Nevertheless, all orbits in the bundle do cross the reference subspace $q^{(1)}=0$ at some moment of time (which may be greater or less than $T_\mathrm{end}$, hence we run the simulation for a slightly longer time to ensure the crossing of this subspace). We determine the time of crossing and the corresponding coordinates for each $k$-th orbit in the following way. In the $N$-body simulation, we store the position and velocity of the Sgr remnant's centre at each timestep $\boldsymbol w_k(t)$ and linearly transform it to $\boldsymbol q_k(t)$. These values are still subject to numerical noise, so we first locate the snapshot closest to $q_k^{(1)}(t)=0$, and then fit a smooth orbit to the centre positions over the interval $\pm 0.1$~Gyr around this time. Finally, we use the fitted smooth trajectory to determine more accurately the time of crossing the reference subspace $T_{\mathrm{cross},k}$ and the corresponding 5d coordinates $q_{\mathrm{cross},k}^{(2-6)}$.

The key point now is that instead of using the linearly transformed initial and end states $\boldsymbol p_{\mathrm{in},k},\; \boldsymbol q_{\mathrm{end},k}$ in the Jacobian (which, of course, would not change its condition number), we define the end state $\tilde{\boldsymbol q}_{\mathrm{end},k}$ of each orbit by the time of crossing and the corresponding 5d coordinates. Of course, the correspondence between $\boldsymbol q_{\mathrm{end},k}$ and $\tilde{\boldsymbol q}_{\mathrm{end},k} \equiv \{T_{\mathrm{cross},k}, q_{\mathrm{cross},k}^{(2-6)}\}$ is well defined, however, this transformation is nonlinear in a favourable way. The main nonlinear effect is the motion along a curved trajectory, which can be followed explicitly, e.g., by numerically integrating a test-particle trajectory in the static MW potential (over such short timescales the actual trajectory of the Sgr remnant's centre in the simulation is well approximated by a test-particle orbit). Similarly, the initial states of actual orbits are already confined to a 5d subspace, but for any point outside this subspace, it can be projected back by following a curvilinear trajectory until crossing $p^{(1)}=0$, thus defining the nonlinearly transformed vector $\tilde{\boldsymbol p}$ that contains the ``flight time'' in the first component and the 5d coordinates of the crossing point in the remaining components. 
Once the curved orbital motion is compensated, the Jacobian of transformation between the initial $\tilde{\boldsymbol p}_{\mathrm{in},k}$ and the final $\tilde{\boldsymbol q}_{\mathrm{end},k}$ states is much better behaved, with the condition number $\lesssim 10$.

To summarize, the Gauss--Newton iterative procedure is applied to the nonlinearly transformed coordinates associated with the initial and final states. The transformation between these coordinates and the actual position and velocity vectors is performed by with numerical integration of a test-particle trajectory. This approach allows us to lead the simulation to the given final state with an accuracy better than $0.05$~kpc and $0.5$~\kms in $3-4$ iterations. We find that this level of accuracy is necessary for an adequate comparison between different simulations (i.e., if the errors in the final position or velocity are larger, the projected kinematic maps are noticeably different).

\begin{table}
\caption{Measurements of the mean PM and its dispersion in 200 Voronoi bins, each bin containing $\sim10^3$ stars.
$\alpha$ and $\delta$ are the average coordinates of stars in each bin; $\overline\mu_\alpha$ and $\overline\mu_\alpha$ are the mean PM components, $\sigma_\alpha$ and $\sigma_\delta$ are the dispersions, and $\rho$ is the correlation coefficient, representing the non-diagonal element of the 2d dispersion tensor.
Statistical uncertainty on PM dispersion is $\lesssim 0.01$~\masyr, and systematic error in mean PM is $\lesssim 0.05$~\masyr.
} \label{tab:pm}
\begin{tabular}{rrrrrrr}
$\alpha\;\;\;\;\;$ & $\delta\;\;\;\;\;$ & $\overline{\mu}_\alpha\;\;$ & $\overline{\mu}_\delta\;\;$ & $\sigma_\alpha\;\;$ & $\sigma_\delta\;\;$ & $\rho\;\;$ \\
\multicolumn{2}{c}{\;\footnotesize deg} & \multicolumn{2}{c}{\footnotesize\masyr} & \multicolumn{2}{c}{\footnotesize\masyr} & \\
\hline
283.765 & -30.485 & -2.729 & -1.364 & 0.107 & 0.091 & 0.294 \\
284.094 & -30.513 & -2.726 & -1.355 & 0.135 & 0.119 & 0.422 \\
283.782 & -30.812 & -2.702 & -1.370 & 0.119 & 0.113 & 0.359 \\
283.434 & -30.604 & -2.702 & -1.354 & 0.118 & 0.102 & 0.188 \\
283.235 & -30.937 & -2.716 & -1.346 & 0.154 & 0.133 & 0.314 \\
283.054 & -30.373 & -2.669 & -1.341 & 0.144 & 0.124 & 0.381 \\
283.524 & -30.254 & -2.713 & -1.352 & 0.115 & 0.103 & 0.197 \\
282.764 & -30.633 & -2.672 & -1.309 & 0.158 & 0.127 & 0.281 \\
282.746 & -31.023 & -2.675 & -1.312 & 0.126 & 0.126 & 0.389 \\
283.069 & -30.015 & -2.712 & -1.347 & 0.138 & 0.135 & 0.327 \\
283.933 & -30.212 & -2.710 & -1.365 & 0.138 & 0.114 & 0.310 \\
283.303 & -29.670 & -2.722 & -1.352 & 0.152 & 0.149 & 0.264 \\
283.710 & -29.837 & -2.690 & -1.360 & 0.146 & 0.128 & 0.269 \\
282.517 & -30.270 & -2.655 & -1.330 & 0.149 & 0.123 & 0.174 \\
282.582 & -29.869 & -2.698 & -1.337 & 0.120 & 0.137 & 0.378 \\
282.172 & -30.879 & -2.691 & -1.342 & 0.132 & 0.111 & 0.425 \\
282.105 & -30.494 & -2.693 & -1.352 & 0.125 & 0.123 & 0.235 \\
282.054 & -30.063 & -2.729 & -1.324 & 0.145 & 0.116 & 0.236 \\
282.576 & -29.494 & -2.722 & -1.315 & 0.151 & 0.109 & 0.398 \\
282.049 & -29.667 & -2.705 & -1.326 & 0.127 & 0.126 & 0.299 \\
283.102 & -29.296 & -2.704 & -1.366 & 0.153 & 0.124 & 0.339 \\
283.576 & -28.994 & -2.715 & -1.380 & 0.147 & 0.143 & 0.411 \\
284.165 & -29.838 & -2.724 & -1.384 & 0.135 & 0.092 & 0.092 \\
283.984 & -29.338 & -2.715 & -1.364 & 0.116 & 0.124 & 0.376 \\
282.680 & -28.973 & -2.711 & -1.334 & 0.162 & 0.130 & 0.408 \\
282.752 & -28.253 & -2.702 & -1.355 & 0.133 & 0.119 & 0.355 \\
281.926 & -29.288 & -2.720 & -1.329 & 0.143 & 0.123 & 0.301 \\
281.562 & -29.904 & -2.720 & -1.329 & 0.156 & 0.132 & 0.111 \\
282.087 & -28.847 & -2.736 & -1.335 & 0.134 & 0.132 & 0.320 \\
281.612 & -30.400 & -2.698 & -1.327 & 0.145 & 0.116 & 0.403 \\
281.390 & -29.360 & -2.710 & -1.317 & 0.133 & 0.134 & 0.519 \\
283.442 & -28.286 & -2.699 & -1.380 & 0.157 & 0.143 & 0.224 \\
282.029 & -28.061 & -2.706 & -1.336 & 0.147 & 0.114 & 0.195 \\
282.731 & -26.889 & -2.688 & -1.396 & 0.138 & 0.142 & 0.557 \\
281.473 & -28.744 & -2.686 & -1.312 & 0.132 & 0.137 & 0.403 \\
281.043 & -29.782 & -2.692 & -1.321 & 0.141 & 0.124 & 0.322 \\
281.565 & -30.975 & -2.707 & -1.326 & 0.151 & 0.135 & 0.416 \\
281.018 & -30.256 & -2.676 & -1.309 & 0.140 & 0.120 & 0.356 \\
280.922 & -29.080 & -2.704 & -1.318 & 0.168 & 0.124 & 0.156 \\
281.011 & -30.721 & -2.690 & -1.295 & 0.139 & 0.121 & 0.264 \\
280.535 & -29.582 & -2.693 & -1.304 & 0.149 & 0.135 & 0.299 \\
281.315 & -27.971 & -2.714 & -1.325 & 0.148 & 0.128 & 0.192 \\
280.741 & -28.573 & -2.704 & -1.334 & 0.169 & 0.133 & 0.465 \\
280.418 & -27.949 & -2.724 & -1.313 & 0.141 & 0.108 & 0.213 \\
280.239 & -29.071 & -2.725 & -1.327 & 0.160 & 0.128 & 0.393 \\
285.259 & -28.213 & -2.680 & -1.420 & 0.151 & 0.141 & 0.604 \\
284.199 & -28.309 & -2.684 & -1.397 & 0.126 & 0.130 & 0.338 \\
284.539 & -28.871 & -2.690 & -1.392 & 0.156 & 0.133 & 0.566 \\
280.658 & -26.746 & -2.722 & -1.343 & 0.115 & 0.117 & 0.363 \\
279.825 & -28.551 & -2.763 & -1.302 & 0.161 & 0.154 & 0.405 \\
279.585 & -27.804 & -2.717 & -1.285 & 0.158 & 0.113 & 0.329 \\
279.959 & -29.724 & -2.731 & -1.265 & 0.143 & 0.127 & 0.242 \\
280.445 & -30.228 & -2.688 & -1.298 & 0.165 & 0.137 & 0.241 \\
279.509 & -29.066 & -2.759 & -1.305 & 0.158 & 0.118 & 0.270 \\
280.339 & -30.857 & -2.678 & -1.289 & 0.132 & 0.123 & 0.305 \\
278.895 & -28.174 & -2.714 & -1.286 & 0.155 & 0.115 & 0.393 \\
\hline
\end{tabular}
\end{table}
\begin{table}
\contcaption{Mean PM and its dispersion}
\begin{tabular}{rrrrrrr}
$\alpha\;\;\;\;\;$ & $\delta\;\;\;\;\;$ & $\overline{\mu}_\alpha\;\;$ & $\overline{\mu}_\delta\;\;$ & $\sigma_\alpha\;\;$ & $\sigma_\delta\;\;$ & $\rho\;\;$ \\
\hline
279.294 & -29.669 & -2.718 & -1.268 & 0.150 & 0.135 & 0.429 \\
280.850 & -31.328 & -2.686 & -1.267 & 0.146 & 0.127 & 0.420 \\
281.610 & -31.526 & -2.713 & -1.300 & 0.148 & 0.110 & 0.453 \\
282.285 & -31.326 & -2.667 & -1.300 & 0.134 & 0.122 & 0.376 \\
284.565 & -29.675 & -2.706 & -1.382 & 0.143 & 0.122 & 0.463 \\
279.733 & -30.279 & -2.698 & -1.265 & 0.153 & 0.140 & 0.341 \\
279.998 & -31.736 & -2.692 & -1.256 & 0.154 & 0.146 & 0.557 \\
279.619 & -30.875 & -2.683 & -1.272 & 0.143 & 0.140 & 0.481 \\
278.813 & -28.837 & -2.751 & -1.279 & 0.172 & 0.144 & 0.584 \\
278.903 & -30.276 & -2.716 & -1.245 & 0.151 & 0.154 & 0.459 \\
278.639 & -29.457 & -2.697 & -1.220 & 0.144 & 0.141 & 0.605 \\
278.616 & -26.675 & -2.721 & -1.281 & 0.146 & 0.098 & 0.244 \\
277.989 & -27.948 & -2.690 & -1.260 & 0.157 & 0.109 & 0.256 \\
286.233 & -28.129 & -2.661 & -1.461 & 0.148 & 0.137 & 0.615 \\
282.825 & -31.578 & -2.711 & -1.338 & 0.140 & 0.131 & 0.311 \\
284.903 & -29.292 & -2.673 & -1.378 & 0.118 & 0.126 & 0.413 \\
285.584 & -29.011 & -2.676 & -1.425 & 0.147 & 0.132 & 0.527 \\
282.199 & -31.868 & -2.703 & -1.286 & 0.144 & 0.143 & 0.339 \\
278.262 & -30.851 & -2.695 & -1.219 & 0.195 & 0.136 & 0.536 \\
280.916 & -32.148 & -2.733 & -1.294 & 0.132 & 0.131 & 0.203 \\
278.042 & -29.884 & -2.703 & -1.229 & 0.194 & 0.123 & 0.230 \\
277.970 & -29.002 & -2.709 & -1.240 & 0.152 & 0.135 & 0.380 \\
276.844 & -26.823 & -2.766 & -1.247 & 0.139 & 0.106 & 0.425 \\
278.896 & -31.465 & -2.741 & -1.232 & 0.150 & 0.138 & 0.646 \\
281.804 & -32.457 & -2.712 & -1.289 & 0.132 & 0.133 & 0.381 \\
278.741 & -32.807 & -2.738 & -1.197 & 0.162 & 0.145 & 0.373 \\
283.333 & -31.321 & -2.695 & -1.340 & 0.124 & 0.120 & 0.346 \\
283.865 & -31.289 & -2.671 & -1.327 & 0.137 & 0.125 & 0.418 \\
284.189 & -30.956 & -2.683 & -1.347 & 0.133 & 0.118 & 0.414 \\
283.147 & -32.066 & -2.688 & -1.341 & 0.138 & 0.138 & 0.143 \\
284.749 & -30.094 & -2.695 & -1.380 & 0.112 & 0.140 & 0.300 \\
284.487 & -30.350 & -2.665 & -1.372 & 0.135 & 0.115 & 0.256 \\
277.216 & -28.457 & -2.705 & -1.238 & 0.167 & 0.141 & 0.353 \\
276.853 & -29.458 & -2.706 & -1.194 & 0.161 & 0.111 & 0.214 \\
280.663 & -33.480 & -2.712 & -1.218 & 0.154 & 0.120 & 0.397 \\
284.374 & -31.366 & -2.678 & -1.388 & 0.126 & 0.100 & 0.116 \\
277.206 & -30.385 & -2.706 & -1.180 & 0.189 & 0.124 & 0.536 \\
276.969 & -31.963 & -2.743 & -1.166 & 0.152 & 0.112 & 0.376 \\
284.537 & -30.735 & -2.691 & -1.359 & 0.150 & 0.144 & \!\!-0.014 \\
282.656 & -32.799 & -2.695 & -1.320 & 0.137 & 0.144 & 0.390 \\
283.733 & -31.741 & -2.664 & -1.349 & 0.158 & 0.121 & 0.447 \\
284.022 & -32.244 & -2.686 & -1.344 & 0.130 & 0.111 & 0.186 \\
285.191 & -29.952 & -2.708 & -1.396 & 0.130 & 0.145 & 0.324 \\
284.424 & -31.862 & -2.685 & -1.369 & 0.131 & 0.132 & 0.308 \\
285.020 & -30.588 & -2.699 & -1.386 & 0.133 & 0.103 & 0.336 \\
284.760 & -31.064 & -2.733 & -1.370 & 0.160 & 0.132 & 0.493 \\
284.970 & -31.419 & -2.705 & -1.373 & 0.141 & 0.126 & 0.230 \\
283.771 & -32.936 & -2.689 & -1.323 & 0.157 & 0.143 & 0.494 \\
284.681 & -32.698 & -2.672 & -1.334 & 0.144 & 0.133 & 0.323 \\
285.227 & -30.991 & -2.690 & -1.368 & 0.146 & 0.137 & 0.155 \\
285.452 & -30.452 & -2.679 & -1.381 & 0.118 & 0.121 & 0.428 \\
285.568 & -29.674 & -2.702 & -1.419 & 0.141 & 0.131 & 0.319 \\
285.009 & -31.917 & -2.688 & -1.400 & 0.120 & 0.124 & 0.389 \\
275.030 & -30.718 & -2.718 & -1.117 & 0.157 & 0.100 & 0.418 \\
275.705 & -28.477 & -2.695 & -1.171 & 0.150 & 0.093 & 0.368 \\
284.736 & -26.994 & -2.688 & -1.429 & 0.135 & 0.115 & 0.516 \\
286.164 & -29.360 & -2.662 & -1.437 & 0.163 & 0.121 & 0.713 \\
285.766 & -30.813 & -2.693 & -1.411 & 0.131 & 0.140 & 0.372 \\
285.836 & -30.186 & -2.694 & -1.397 & 0.126 & 0.122 & 0.349 \\
285.695 & -31.233 & -2.649 & -1.384 & 0.139 & 0.123 & 0.372 \\
285.555 & -31.646 & -2.682 & -1.364 & 0.153 & 0.123 & 0.358 \\
286.398 & -30.058 & -2.676 & -1.426 & 0.128 & 0.115 & 0.432 \\
285.382 & -32.409 & -2.638 & -1.358 & 0.149 & 0.147 & 0.236 \\
286.322 & -30.553 & -2.650 & -1.416 & 0.153 & 0.134 & 0.310 \\
\hline
\end{tabular}
\end{table}
\begin{table}
\contcaption{Mean PM and its dispersion}
\begin{tabular}{rrrrrrr}
$\alpha\;\;\;\;\;$ & $\delta\;\;\;\;\;$ & $\overline{\mu}_\alpha\;\;$ & $\overline{\mu}_\delta\;\;$ & $\sigma_\alpha\;\;$ & $\sigma_\delta\;\;$ & $\rho\;\;$ \\
\hline
286.291 & -31.080 & -2.676 & -1.383 & 0.135 & 0.128 & 0.394 \\
286.782 & -29.558 & -2.678 & -1.457 & 0.141 & 0.146 & 0.369 \\
287.304 & -28.657 & -2.678 & -1.480 & 0.152 & 0.141 & 0.578 \\
286.092 & -31.620 & -2.673 & -1.385 & 0.143 & 0.133 & 0.306 \\
285.918 & -32.234 & -2.690 & -1.381 & 0.148 & 0.140 & 0.461 \\
286.023 & -33.298 & -2.625 & -1.360 & 0.143 & 0.130 & 0.351 \\
286.961 & -30.588 & -2.662 & -1.403 & 0.150 & 0.123 & 0.554 \\
287.535 & -29.476 & -2.658 & -1.437 & 0.139 & 0.148 & 0.485 \\
286.646 & -31.510 & -2.680 & -1.423 & 0.133 & 0.131 & 0.439 \\
286.518 & -32.036 & -2.692 & -1.412 & 0.156 & 0.137 & 0.160 \\
288.447 & -29.459 & -2.669 & -1.504 & 0.151 & 0.146 & 0.424 \\
286.916 & -31.037 & -2.688 & -1.434 & 0.160 & 0.133 & 0.053 \\
287.200 & -31.489 & -2.682 & -1.486 & 0.154 & 0.153 & 0.350 \\
286.502 & -32.742 & -2.640 & -1.335 & 0.160 & 0.139 & 0.450 \\
287.641 & -30.613 & -2.687 & -1.491 & 0.168 & 0.122 & 0.432 \\
287.234 & -30.113 & -2.640 & -1.424 & 0.169 & 0.158 & 0.508 \\
287.084 & -32.317 & -2.598 & -1.372 & 0.148 & 0.131 & 0.279 \\
284.963 & -33.498 & -2.669 & -1.355 & 0.149 & 0.131 & 0.416 \\
287.491 & -31.917 & -2.592 & -1.364 & 0.137 & 0.135 & 0.420 \\
288.700 & -28.522 & -2.634 & -1.488 & 0.132 & 0.135 & 0.573 \\
287.785 & -31.061 & -2.654 & -1.427 & 0.145 & 0.138 & 0.483 \\
288.005 & -31.524 & -2.594 & -1.364 & 0.158 & 0.128 & 0.164 \\
287.274 & -33.072 & -2.644 & -1.349 & 0.157 & 0.144 & 0.287 \\
287.874 & -32.550 & -2.633 & -1.401 & 0.163 & 0.109 & 0.486 \\
288.730 & -30.502 & -2.614 & -1.413 & 0.177 & 0.156 & 0.514 \\
283.113 & -34.182 & -2.701 & -1.285 & 0.152 & 0.139 & 0.393 \\
288.114 & -30.148 & -2.658 & -1.462 & 0.145 & 0.127 & 0.524 \\
288.365 & -32.075 & -2.640 & -1.427 & 0.145 & 0.137 & 0.417 \\
287.822 & -27.002 & -2.686 & -1.540 & 0.152 & 0.133 & 0.346 \\
288.594 & -31.115 & -2.593 & -1.393 & 0.167 & 0.129 & 0.405 \\
288.907 & -31.671 & -2.676 & -1.474 & 0.171 & 0.139 & 0.361 \\
289.361 & -30.010 & -2.604 & -1.446 & 0.173 & 0.139 & 0.477 \\
288.434 & -33.379 & -2.648 & -1.432 & 0.164 & 0.123 & 0.274 \\
287.424 & -33.928 & -2.652 & -1.389 & 0.157 & 0.136 & 0.345 \\
288.892 & -32.669 & -2.642 & -1.455 & 0.175 & 0.122 & 0.373 \\
289.512 & -31.043 & -2.665 & -1.476 & 0.164 & 0.122 & 0.485 \\
290.381 & -29.713 & -2.630 & -1.548 & 0.172 & 0.132 & 0.478 \\
289.575 & -31.952 & -2.652 & -1.469 & 0.161 & 0.128 & 0.378 \\
290.404 & -30.692 & -2.647 & -1.514 & 0.172 & 0.138 & 0.574 \\
289.935 & -33.470 & -2.605 & -1.436 & 0.151 & 0.124 & 0.441 \\
289.869 & -32.681 & -2.645 & -1.432 & 0.156 & 0.129 & 0.413 \\
286.103 & -34.866 & -2.638 & -1.315 & 0.159 & 0.132 & 0.410 \\
290.331 & -31.558 & -2.650 & -1.508 & 0.191 & 0.146 & 0.593 \\
290.677 & -32.224 & -2.649 & -1.475 & 0.175 & 0.133 & 0.421 \\
291.573 & -30.381 & -2.607 & -1.587 & 0.197 & 0.164 & 0.560 \\
294.703 & -27.799 & -2.639 & -1.747 & 0.184 & 0.165 & 0.595 \\
291.225 & -31.366 & -2.674 & -1.538 & 0.179 & 0.145 & 0.316 \\
291.173 & -33.017 & -2.637 & -1.490 & 0.178 & 0.137 & 0.404 \\
288.993 & -34.191 & -2.649 & -1.421 & 0.158 & 0.133 & 0.341 \\
291.786 & -32.263 & -2.626 & -1.528 & 0.192 & 0.140 & 0.395 \\
293.116 & -30.075 & -2.630 & -1.642 & 0.214 & 0.151 & 0.625 \\
294.924 & -29.983 & -2.588 & -1.663 & 0.210 & 0.167 & 0.584 \\
291.000 & -34.156 & -2.610 & -1.446 & 0.166 & 0.133 & 0.336 \\
292.354 & -31.342 & -2.649 & -1.585 & 0.202 & 0.131 & 0.498 \\
292.552 & -33.304 & -2.605 & -1.511 & 0.186 & 0.139 & 0.460 \\
289.814 & -35.480 & -2.676 & -1.399 & 0.199 & 0.137 & 0.468 \\
292.343 & -34.463 & -2.605 & -1.470 & 0.173 & 0.117 & 0.416 \\
295.329 & -31.994 & -2.619 & -1.649 & 0.199 & 0.150 & 0.542 \\
292.984 & -32.381 & -2.611 & -1.554 & 0.198 & 0.132 & 0.354 \\
293.936 & -34.158 & -2.602 & -1.528 & 0.185 & 0.122 & 0.516 \\
294.145 & -32.834 & -2.609 & -1.567 & 0.224 & 0.135 & 0.458 \\
296.891 & -31.147 & -2.597 & -1.717 & 0.248 & 0.173 & 0.651 \\
295.535 & -33.759 & -2.571 & -1.593 & 0.213 & 0.131 & 0.520 \\
297.827 & -28.838 & -2.617 & -1.824 & 0.208 & 0.148 & 0.436 \\
\hline
\end{tabular}
\end{table}
\begin{table}
\contcaption{Mean PM and its dispersion}
\begin{tabular}{rrrrrrr}
$\alpha\;\;\;\;\;$ & $\delta\;\;\;\;\;$ & $\overline{\mu}_\alpha\;\;$ & $\overline{\mu}_\delta\;\;$ & $\sigma_\alpha\;\;$ & $\sigma_\delta\;\;$ & $\rho\;\;$ \\
\hline
297.095 & -32.962 & -2.586 & -1.679 & 0.252 & 0.151 & 0.650 \\
289.794 & -28.850 & -2.615 & -1.508 & 0.137 & 0.143 & 0.462 \\
291.297 & -27.500 & -2.606 & -1.587 & 0.148 & 0.141 & 0.492 \\
291.986 & -29.258 & -2.663 & -1.624 & 0.175 & 0.163 & 0.546 \\
293.817 & -31.333 & -2.598 & -1.619 & 0.189 & 0.146 & 0.481 \\
294.212 & -35.944 & -2.631 & -1.493 & 0.217 & 0.123 & 0.392 \\
296.620 & -35.052 & -2.628 & -1.627 & 0.250 & 0.146 & 0.518 \\
298.754 & -33.917 & -2.610 & -1.712 & 0.255 & 0.160 & 0.582 \\
299.136 & -31.828 & -2.613 & -1.787 & 0.325 & 0.164 & 0.658 \\
300.761 & -29.470 & -2.632 & -1.917 & 0.310 & 0.206 & 0.730 \\
301.738 & -32.207 & -2.645 & -1.861 & 0.332 & 0.212 & 0.743 \\
301.394 & -34.549 & -2.636 & -1.803 & 0.293 & 0.165 & 0.676 \\
298.987 & -36.214 & -2.630 & -1.634 & 0.240 & 0.148 & 0.607 \\
305.239 & -30.134 & -2.726 & -2.122 & 0.356 & 0.239 & 0.699 \\
305.966 & -33.243 & -2.759 & -2.054 & 0.368 & 0.235 & 0.676 \\
304.578 & -35.981 & -2.731 & -1.879 & 0.286 & 0.170 & 0.695 \\
\hline
\end{tabular}
\end{table}
\begin{table}\caption{Measurements of the line-of-sight velocity and its dispersion in 36 Voronoi bins.
$\alpha$ and $\delta$ are the average coordinates of stars in each bin; $\overline{v}_\mathrm{los}$ is the mean heliocentric line-of-sight velocity, $\sigma_\mathrm{los}$ is its dispersion, and $\epsilon_v, \epsilon_\sigma$ are their statistical uncertainties. The final column is the number of stars in the bin.
} \label{tab:vlos}
\begin{tabular}{rrrrrrr}
$\alpha\;\;\;\;\;$ & $\delta\;\;\;\;\;$ & $\overline{v}_\mathrm{los}\;$ & $\sigma_\mathrm{los}\;$ & $\epsilon_v\;$ & $\epsilon_\sigma\;$ & $N_\star$\\
\multicolumn{2}{c}{\;\footnotesize deg} & \multicolumn{2}{c}{\footnotesize\kms} & \multicolumn{2}{c}{\footnotesize\kms} & \\
\hline
287.570 & -32.110 & 143.09 & 13.72 & 1.30 & 0.96 & 115 \\
285.814 & -30.671 & 140.50 & 13.55 & 1.10 & 0.79 & 155 \\
285.764 & -32.520 & 148.96 & 12.85 & 1.49 & 1.06 & 75 \\
280.350 & -28.873 & 144.20 & 13.83 & 1.47 & 1.05 & 89 \\
279.105 & -29.146 & 147.20 & 14.16 & 1.68 & 1.21 & 73 \\
279.780 & -32.368 & 152.03 & 9.53 & 2.80 & 2.07 & 14 \\
281.134 & -27.174 & 139.75 & 14.29 & 3.88 & 2.77 & 14 \\
280.956 & -29.755 & 146.36 & 13.35 & 1.28 & 0.91 & 110 \\
282.744 & -32.031 & 152.40 & 14.66 & 1.41 & 1.01 & 109 \\
282.813 & -28.729 & 137.52 & 12.27 & 1.52 & 1.10 & 67 \\
282.885 & -32.910 & 157.61 & 16.95 & 2.18 & 1.55 & 61 \\
283.922 & -32.266 & 153.91 & 12.52 & 1.36 & 0.97 & 86 \\
283.526 & -31.431 & 149.21 & 13.44 & 1.09 & 0.78 & 155 \\
283.853 & -30.562 & 143.36 & 10.58 & 0.66 & 0.47 & 266 \\
284.036 & -29.705 & 141.98 & 13.57 & 1.46 & 1.06 & 88 \\
284.425 & -27.626 & 125.63 & 21.80 & 4.09 & 2.93 & 29 \\
284.675 & -33.000 & 149.08 & 14.89 & 1.81 & 1.29 & 68 \\
285.838 & -29.303 & 131.04 & 14.12 & 2.32 & 1.70 & 40 \\
287.059 & -31.233 & 141.20 & 12.60 & 1.19 & 0.85 & 114 \\
285.913 & -33.582 & 146.87 & 14.42 & 2.00 & 1.43 & 53 \\
288.136 & -31.397 & 137.97 & 11.33 & 0.90 & 0.65 & 163 \\
289.395 & -34.525 & 147.11 & 13.46 & 2.67 & 1.91 & 26 \\
290.655 & -31.837 & 134.79 & 12.38 & 1.56 & 1.14 & 67 \\
292.827 & -32.504 & 134.58 & 10.60 & 1.60 & 1.17 & 48 \\
295.205 & -33.004 & 129.66 & 12.99 & 2.65 & 1.90 & 25 \\
298.174 & -33.755 & 128.07 & 11.10 & 2.87 & 2.10 & 16 \\
279.664 & -29.730 & 148.10 & 13.71 & 1.43 & 1.02 & 94 \\
281.834 & -30.273 & 146.12 & 12.01 & 0.93 & 0.66 & 170 \\
281.989 & -31.412 & 151.26 & 13.04 & 1.41 & 1.01 & 88 \\
282.100 & -29.527 & 143.89 & 11.70 & 1.03 & 0.73 & 131 \\
283.024 & -30.355 & 142.41 & 10.31 & 0.98 & 0.69 & 113 \\
284.412 & -28.506 & 133.44 & 14.61 & 2.06 & 1.46 & 51 \\
284.765 & -31.086 & 143.15 & 13.16 & 0.89 & 0.63 & 220 \\
284.922 & -32.070 & 147.81 & 12.83 & 1.04 & 0.74 & 155 \\
285.904 & -31.399 & 142.47 & 12.45 & 1.10 & 0.78 & 131 \\
294.557 & -30.610 & 125.41 & 10.79 & 2.50 & 1.78 & 19 \\
\hline
\end{tabular}
\end{table}


\begin{figure*}
\includegraphics{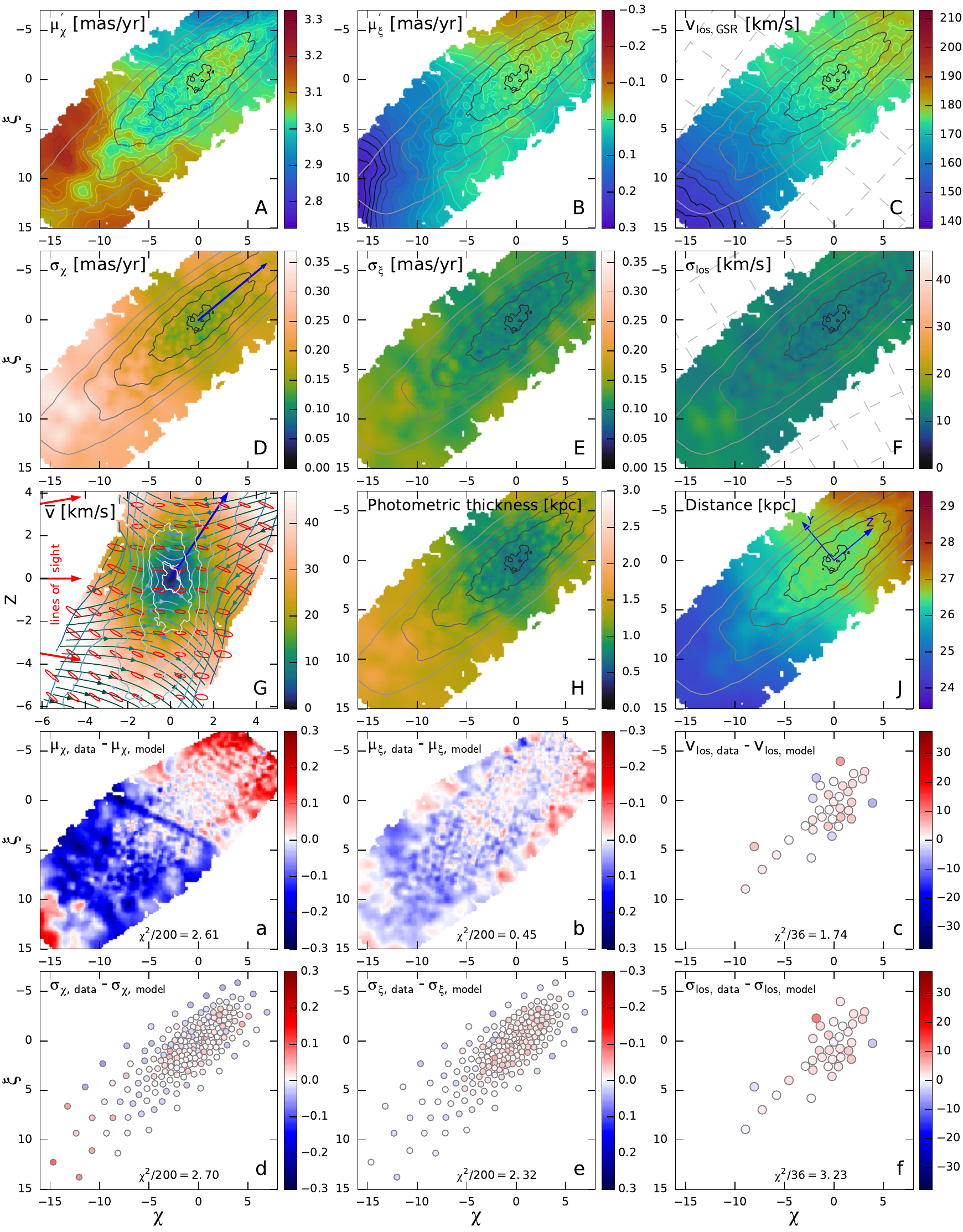}
\caption{Same as Figure~\ref{fig:maps_model}, but for a model that is too much tidally stretched and is more aligned with the stream (Panel G showing the side-on view, as in Figure~\ref{fig:maps_internal}). The trailing side is too close (Panel J), causing a serious misfit in $\mu_\chi'$ (Panel A).
\vspace*{-10pt}}  \label{fig:maps_model1}
\end{figure*}

\begin{figure*}
\includegraphics{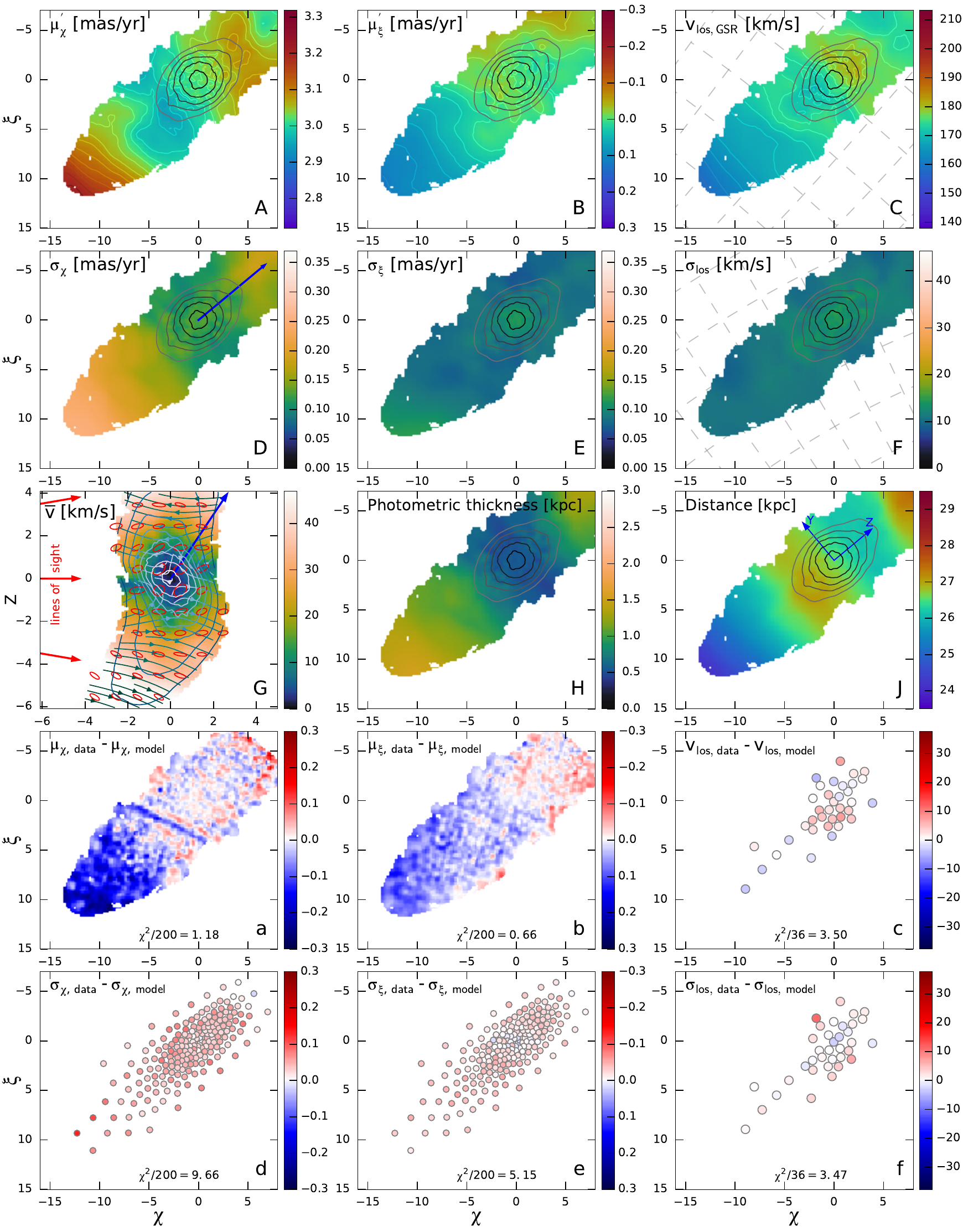}
\caption{Same as Figure~\ref{fig:maps_model}, but for the \citet{Law2010} model (resimulated with a slightly different orbit to better match the present-day position and velocity of Sgr), which is too compact and transitions to the stream too early (Panels A, G).
\vspace*{-10pt}}  \label{fig:maps_model2}
\end{figure*}

\begin{figure*}
\includegraphics{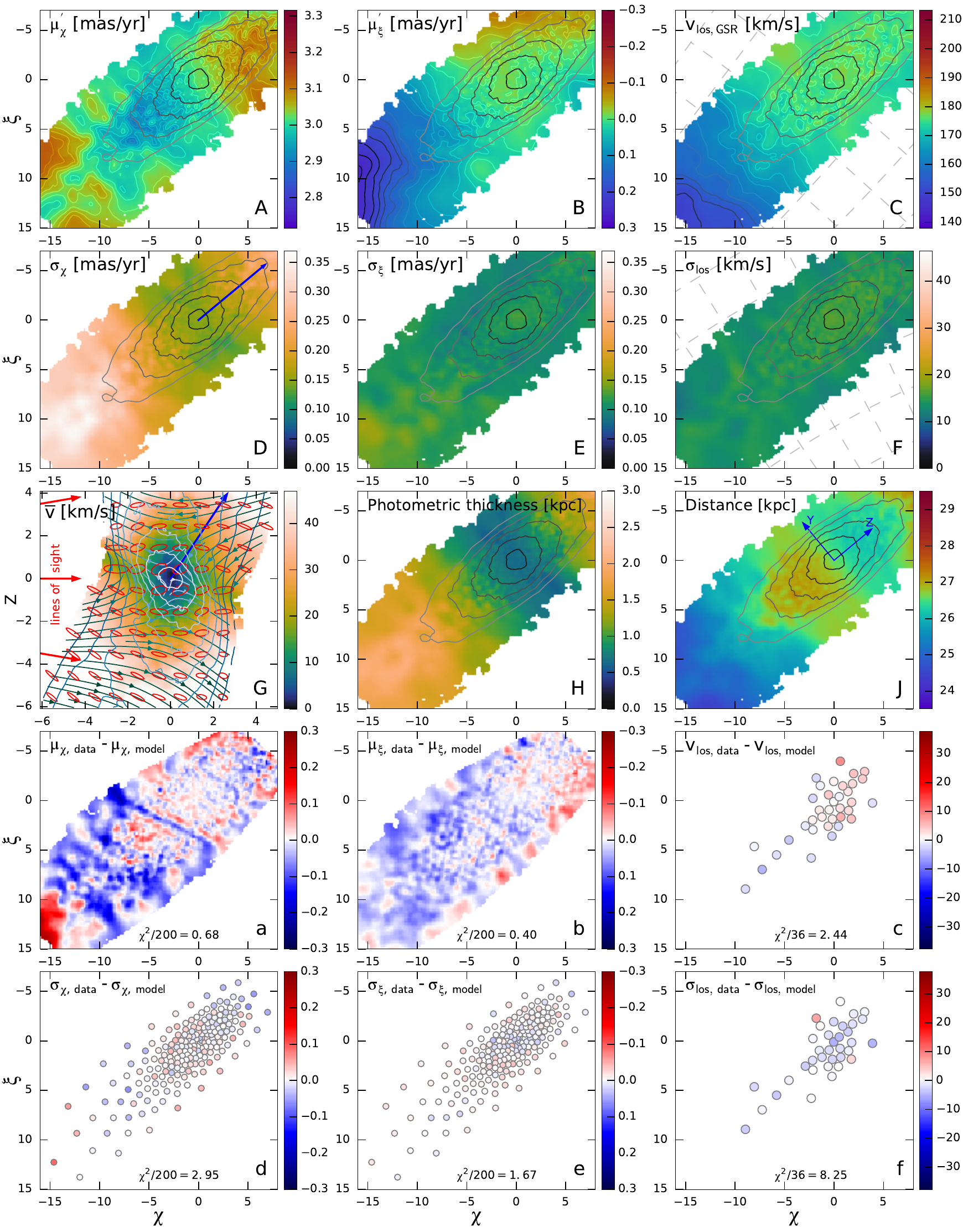}
\caption{Same as Figure~\ref{fig:maps_model}, but a model that has a cuspy dark matter halo. Despite having a mass 1.5$\times$ lower than the fiducial model, its line-of-sight velocity dispersion is still too high (Panel F).
\vspace*{-10pt}}  \label{fig:maps_model3}
\end{figure*}

\begin{figure*}
\includegraphics{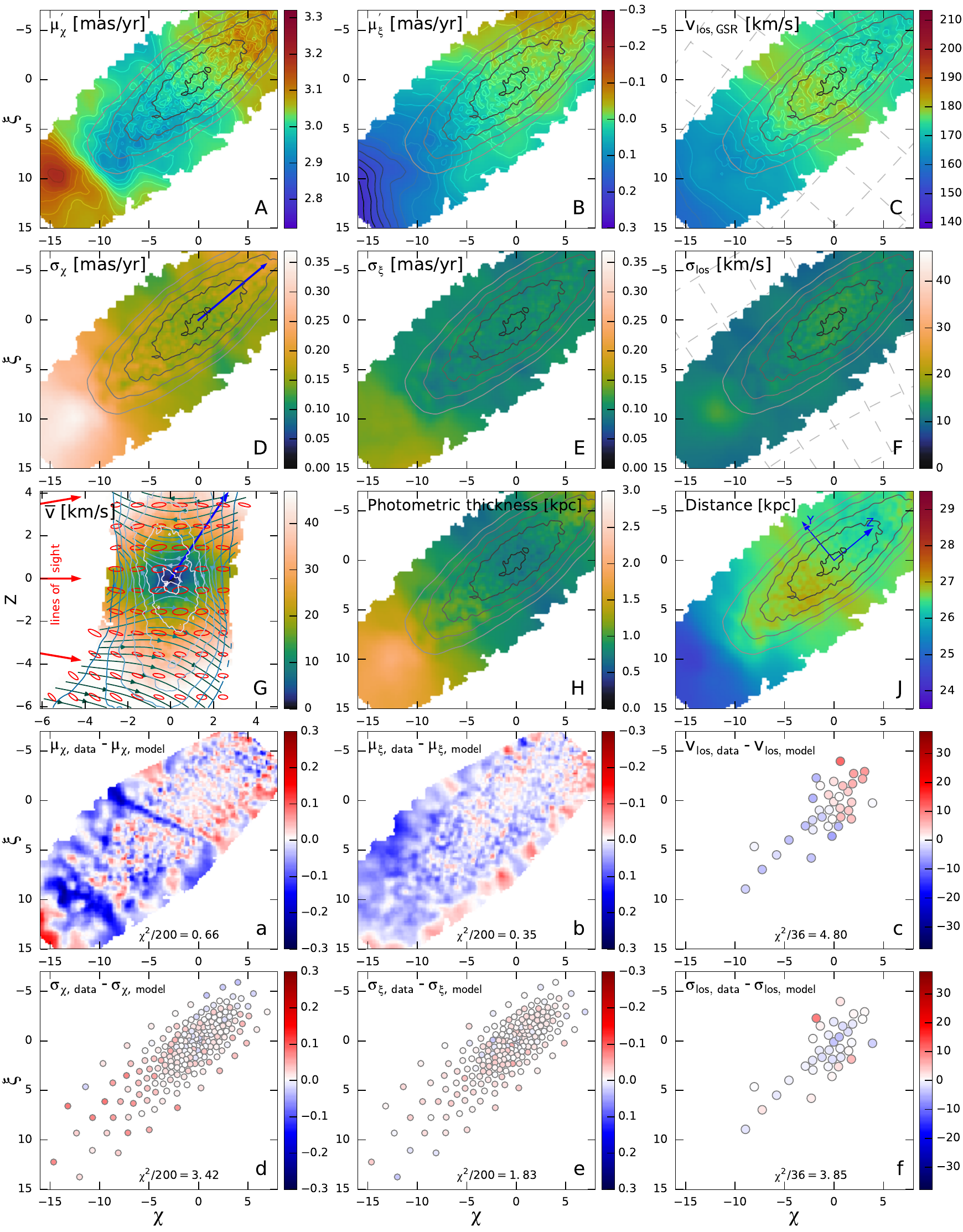}
\caption{Same as Figure~\ref{fig:maps_model}, but a model that is initially rotating and forms a tidally induced bar, as in the \citet{Lokas2010} scenario. The residual rotation is manifested in the non-monotonic $v_\mathrm{los,GSR}$ profile along the major axis (Panel C).
\vspace*{-10pt}}  \label{fig:maps_model4}
\end{figure*}

\end{document}